# Complex-Dynamic Origin of Consciousness and the Critical Choice of Sustainability Transition

ANDREI P. KIRILYUK[*]

Institute of Metal Physics, Kiev, Ukraine

**ABSTRACT.** A quite general interaction process of a multi-component system is analysed by the extended effective potential method liberated from usual limitations of perturbation theory or integrable model. The obtained causally complete solution of the many-body problem reveals the phenomenon of dynamic multivaluedness, or redundance, of emerging, incompatible system realisations and dynamic entanglement of system components within each realisation. The ensuing concept of dynamic complexity (and related intrinsic chaoticity) is absolutely universal and can be applied to the problem of (natural and artificial) intelligence and consciousness that dynamically emerge now as a high enough, properly specified levels of unreduced complexity of a suitable interaction process. Emergent consciousness can be identified with the appearance of bound, permanently localised states in the multivalued brain dynamics from strongly chaotic states of unconscious intelligence, by analogy with classical behaviour emergence from quantum states at the lowest levels of complex world dynamics. We show that the main properties of this dynamically emerging consciousness (and intelligence, at the preceding complexity level) correspond to empirically derived properties of natural consciousness and obtain causally substantiated conclusions about their artificial realisation, including the fundamentally justified paradigm of genuine machine consciousness. This rigorously defined machine consciousness is different from both natural consciousness and any mechanistic, dynamically single-valued imitation of the latter. We use then the same, truly universal concept of complexity to derive equally rigorous conclusions about mental and social implications of this complex-dynamic consciousness concept, demonstrating its critical importance for further progress of science and civilisation.

---

[*] Address for correspondence: Institute of Metal Physics, Solid State Theory Department, 36 Vernadsky Av., Kyiv 03142, Ukraine. E-mail address: Andrei.Kirilyuk@Gmail.com.

# Complex-Dynamic Origin of Consciousness and the Critical Choice of Sustainability Transition

**Back-cover book description.** The problem of rigorous scientific description and understanding of intelligence and consciousness remains unsolved despite its quickly growing importance, including various applications, such as artificial intelligence and machine consciousness. In this book we present a new solution to this problem based on the causally complete solution of the arbitrary many-body interaction problem leading to the universal concept of dynamic complexity and chaoticity in terms of dynamic multivaluedness, or redundance, of explicitly emerging system realisations. Intelligence and consciousness are rigorously specified as certain, high enough levels of this unreduced complexity of natural or artificial brain dynamics, with their observed major properties (escaping otherwise scientifically exact explanation). We show then how the proposed concept of consciousness is closely related to the necessary global sustainability transition specified now as a step-like, unified growth of the level of consciousness and complexity in all human activities and approaches, in science and beyond. The book contains both the rigorous basis and popular discussion oriented to a wide audience of educated readers.

**CONTENTS**

# ABBREVIATIONS
used in the text

EP *for* Effective Potential (introduced in Section 2.1)
SOC *for* Self-Organised Criticality (introduced in Section 2.3)

# 1. Introduction

Canonical science cannot provide the truly scientific, i.e. consistent, fundamental and universal understanding of consciousness considered either as an empirically perceived property of human brain dynamics or a general property of intelligent enough system of any origin (see e.g. [1-12] for various existing approaches to consciousness and further references). Although the problem as such is not new and actually cannot be separated from the "eternal" man's quest for his "ultimate" origin and destination, the last-time development of technology, society and civilisation is quickly changing its status from vague "philosophical" speculations (always remaining with us and inherent to the problem) to much more practically oriented and even critically growing issue with increasingly important consequences at various levels of human activity, from quite new directions of technology development to the deepest changes in individual and social life. Despite visible stagnation (and actual failure) of the previous paradigm of artificial intelligence, largely reduced to various kinds of generally useful, but definitely *non*-intelligent "expert systems" (i.e. actually *false intelligence*), the recently initiated inquiry of *artificial*, or *machine*, *consciousness* [13,14], starts from another, much more constructive level of purposes and demands, involving qualitatively stronger interaction between previously separated disciplines and explicit intention to transcend other barriers of scientific tradition in the direction of *qualitatively* more consistent and rigorous knowledge (this implies, in particular, that machine consciousness should contain at least some *irreducible*, specific properties of *genuine* consciousness). Note that similar to "artificial" intelligence, the "machine" consciousness thesis should certainly include as its integral, starting component the unreduced, provably complete and rigorous understanding of the corresponding *natural* properties of human brain that can then be implemented in artificial systems, with various degree of similarity to the natural prototypes, which should constitute itself an indispensable part (and validity criterion) of the consistent enough theory of conscious intelligence.

The purpose of this work is to present a new, universal theory of consciousness based on the recently elaborated, reality-based concept of dynamic complexity [15-19] and satisfying the mentioned demands of modern technology and social development. Therefore we shall not review the existing other approaches to consciousness and their results (see e.g. [1-14] for some details and further references): it will be enough to note that the latter are reduced to a sufficiently detailed (and certainly indispensable) description of empirically observed aspects and features of consciousness that can serve now as a basis for verification of any proposed "genuine", scientifically rigorous and integrated, understanding of consciousness. Those results will thus be present in our exposure, explicitly and implicitly, in the form of a unified system of correlations between theoretically derived properties and known practical manifestations of consciousness.

Any truly fundamental, "first-principle" and realistic theory of consciousness should acknowledge the *basic* role of *interaction* between the elements of a carrier of consciousness (such as the brain) in its origin and properties. However, the performed extensive study of various oversimplified, perturbative *models* for such interaction is reduced to *substitution* of the real problem solution, showing natural interaction *development*, for its arbitrarily *assumed* (i.e. "guessed" and postulated) result, which can *not* reveal a clear, scientifically exact origin of *emergent*, qualitatively specific properties of intelligence/consciousness, leaving an impression of "something else" being present in the phenomenon of consciousness, a greater "mystery" unifying all its diverse manifestations in a single whole [1,7-11]. We start our analysis by showing that such contradictory situation is due to the fundamentally restricted, always *perturbative* description of interaction process in the conventional science framework and that if one avoids those usual limitations, then many *qualitatively new* properties do emerge simply from the *unreduced solution* of the many-body interaction problem, with direct relevance to both consistent, *universal concept of dynamic complexity* and the phenomenon of consciousness that appears, similar to a more general phenomenon of intelligence, as a *high enough level of the unreduced interaction complexity* [15,16] (i.e. behaviour of a system with a large enough number of connected, strongly interacting elements with the necessary, but not very special, "generic" properties). In particular, we provide (Section 2) the derivation and *mathematically exact* definition of the *universal dynamic complexity* (as well as closely related *chaoticity*), expressed in terms of fundamental *dynamic multivaluedness, or redundance*, and *entanglement* of the unreduced problem solution [15-19], and therefore can proceed with equally rigorous definition of the properties of *intelligence* and *consciousness*, naturally (dynamically) *emerging* in a complicated enough system of interacting elements (Section 3).

We then consider various specific manifestations and "miraculous" parameters of conscious system operation thus obtained, to reveal the consistent system of correlations with the empirically observed features of consciousness (Section 3). Finally, using the same, universally applicable concept of complexity, we rigorously demonstrate the *objective necessity* and thus inevitability of machine consciousness introduction in complicated enough technological and social systems of the modern world, analyse possibilities of practical realisation of truly conscious behaviour in an artificially engineered system, and outline the basic, rigorously substantiated conditions and consequences, both technical and social, of such conscious machinery incorporation into the real technology development (Section 4). We then generalise those conclusions to the entire new, *intrinsically sustainable development paradigm*, showing that it should inevitably start, in the near future, from a qualitatively great, well-specified transition to a higher level of individual and social consciousness called *complexity revolution*, with the only alternative of growing catastrophic degradation in the current, "default" tendency (Section 5).

The obtained results explicitly demonstrate also the possibilities and universality of the unreduced complexity concept, completing the results of its application at lower complexity levels [15,16,20-24] and thus providing an important additional confirmation of its unrestricted consistency, culminating in the *universal law of conservation and transformation, or symmetry, of complexity* [15,16,18]. We show that the unreduced complexity development at superior levels of intelligence and consciousness reproduces some key features and stages of its development at the lowest, quantum and first classical levels of complexity, and this deeply rooted analogy can be quite useful for consistent understanding and efficient use of both lowest and superior levels of complex world dynamics, without any simplified, zero-complexity reduction or fundamentally incorrect, direct mixture between the two, as it happens too often within the fatally reduced framework of canonical, *dynamically single-valued*, or *unitary*, theory (see Section 3 for details and references) .

# 2. Unreduced interaction dynamics and the universal concept of dynamic complexity

## 2.1. Dynamic redundance, emergent chaos and probability

Any system showing the properties of intelligence and consciousness can only be based on the *unrestricted*, autonomous interaction process in a complicated enough network of elements, such as brain neurons or an artificial circuit. We shall now consider such unreduced many-body interaction problem with a generic configuration and demonstrate that its truly "exact", universally nonperturbative solution, liberated from artificial limitations of usual perturbative models, possesses indeed some *qualitatively new* properties that show a unique system of correlations with the observed "miraculous" properties of living, intelligent, and conscious systems, which can *not* be reproduced by any reduced model *in principle*, irrespective of its *technical* sophistication (we shall reveal the exact *fundamental* reason for that).

We take the *real* system configuration in the form of any number of well-specified *elements* (such as natural neurons or their artificial counterparts) with arbitrary interaction between them, so that every element interacts, in principle, with every other one by a known law. The detailed structure of interaction, or system element *connections*, can eventually be specified for each particular case, but we formulate our *universally* applicable, global description in terms of arbitrary interaction process (or "many-body problem"). Each element has its known internal dynamics, universally expressed by its "generalised Hamiltonian" (which is reduced eventually to a complexity measure, see below) and the corresponding "eigen-solutions". Therefore, those separate (non-interacting) system elements usually represent themselves (quasi) integrable, "simple" enough systems, but we do not impose any such formal limitation, assuming instead that each separate element dynamics is known in the form of explicit solution. As follows from empirical data, that individual element dynamics is typically reduced to several discrete, well-defined (stable enough) states of an element, between which it can switch under the influence of external action (where these transitions between internal element states can be accompanied by signals sent to the "outside", i.e. to other elements, thus realising the global interaction process). The problem will consist then in description and under-standing of evolution of the entire system of interacting elements and its emerging properties and structures, in their unreduced, realistic version (i.e. in principle without any "convenient" approximation of a conventional theory that can be convenient for it, but is always fatal for the real system dynamics, where it typically "kills", as we shall see, just the most interesting features). It is important to emphasize here, in accord with the following analysis, that all the specific, "miraculous" features of the global system behaviour, including intelligence and consciousness, originate just in the development of *unreduced global interaction* between elements, rather than in some very special, "tricky" properties of individual elements and interactions (remaining in reality simple enough), as it is inevitably assumed in conventional, perturbative theories.

The unreduced system dynamics is described by the *existence equation* for the system *state function*, $\Psi$, which actually generalises various model equations and can be derived in a self-consistent way as indeed a universal expression of arbitrary system dynamics [15-19] (see also below):

$$\left\{ \sum_{k=0}^{N} \left[ h_k(q_k) + \sum_{l>k}^{N} V_{kl}(q_k, q_l) \right] \right\} \Psi(Q) = E\Psi(Q), \tag{1}$$

where $q_k$ is the degree of freedom of the *k*-th system element, $Q = (q_0, q_1, ..., q_N)$ is the set of all degrees of freedom (both individual and common), $h_k(q_k)$ is the *k*-th element Hamiltonian, *N* is the number of elements, $V_{kl}(q_k, q_l)$ is the potential of interaction between the *k*-th and *l*-th elements, *E* is the eigenvalue of the generalised system Hamiltonian (i.e. "generalised energy" representing a measure of dynamic complexity), and $\Psi(Q)$ is the system state-function describing completely its configuration. Note that the interaction potential can be generalised to any many-element version, while the existence equation (1) actually includes its time-dependent form (obtained by energy replacement by the time derivative operator and considering one of the degrees of freedom as the time variable). It is important also that the starting existing equation, eq. (1), serves merely as concise mathematical expression of the many-body problem and does not contain any simplifying assumption about system dynamics or configuration (in particular, any formal, model "nonlinearity" of the starting equation is compatible with the following analysis). This means also that the elements of the (conscious) system "environment" can be included, if necessary, in the total system composition, described by eq. (1).

It can often be useful to start from another, equivalent form of existence equation, where one of the degrees of freedom, say $q_0 \equiv \xi$, is explicitly separated from all other ones, $\{q_k\} \equiv Q$ $(k = 1, 2, ..., N)$, so that $\xi$ can be interpreted as common, e.g. spatial, system variable(s), characterising its "global" configuration or interaction, while $\{q_k\}$ may describe "internal" degrees of freedom of the corresponding elements:

$$\left\{ h_0(\xi) + \sum_{k=1}^{N} \left[ h_k(q_k) + V_{0k}(\xi, q_k) + \sum_{l>k}^{N} V_{kl}(q_k, q_l) \right] \right\} \Psi(\xi, Q) = E\Psi(\xi, Q). \quad (2)$$

We shall proceed with this form of the existence equation and consider that $1 \leq k, l \leq N$ everywhere below.

Now we can use the known solutions for the free components,

$$h_k(q_k) \varphi_{kn_k}(q_k) = \varepsilon_{n_k} \varphi_{kn_k}(q_k), \quad (3)$$

where $\{\varphi_{kn_k}(q_k)\}$ and $\{\varepsilon_{n_k}\}$ are the eigenfunctions and eigenvalues of the $k$-th component Hamiltonian $h_k(q_k)$, and the eigenfunctions $\{\varphi_{kn_k}(q_k)\}$ form the complete set of orthonormal functions. Expanding the total system state-function $\Psi(q_0, q_1, ..., q_N)$ over complete sets of eigenfunctions $\{\varphi_{kn_k}(q_k)\}$ for the "functional" degrees of freedom $(q_1, ..., q_N) \equiv Q$, we are left with functions depending only on the selected "structural" degrees of freedom $q_0 \equiv \xi$:

$$\Psi(q_0, q_1, ..., q_N) \equiv \Psi(\xi, Q) = \sum_n \psi_n(\xi) \Phi_n(Q), \quad (4)$$

where the summation is performed over all eigenstate combinations $n \equiv (n_1, n_2, ..., n_N)$ and we designated $\Phi_n(Q) \equiv \varphi_{1n}(q_1)\varphi_{2n}(q_2)...\varphi_{Nn}(q_N)$ for brevity. Inserting the expansion of eq. (4) into eq. (2), multiplying by $\Phi_n^*(Q)$ and integrating over all variables $Q$ (using the eigenfunction orthonormality), we get the following system of equations for $\psi_n(\xi)$:

$$H_n(\xi)\psi_n(\xi) + \sum_{n' \neq n} V_{nn'}(\xi)\psi_{n'}(\xi) = \eta_n \psi_n(\xi) - V_{n0}(\xi)\psi_0(\xi), \quad (5)$$

where

$$\eta_n \equiv E - \varepsilon_n, \quad \varepsilon_n \equiv \sum_k \varepsilon_{n_k}, \quad (6)$$

$$H_n(\xi) = h_0(\xi) + V_{nn}(\xi), \quad (7)$$

$$V_{nn'}(\xi) = \sum_k \left[ V_{0k}^{nn'}(\xi) + \sum_{l>k} V_{kl}^{nn'} \right], \qquad (8a)$$

$$V_{0k}^{nn'}(\xi) = \int_{\Omega_Q} dQ \Phi_n^*(Q) V_{0k}(\xi, q_k) \Phi_{n'}(Q), \qquad (8b)$$

$$V_{kl}^{nn'} = \int_{\Omega_Q} dQ \Phi_n^*(Q) V_{kl}(q_k, q_l) \Phi_{n'}(Q). \qquad (8c)$$

It will be convenient to separate the equation for $\psi_0(\xi)$ in the system of equations (5), describing the usually measured generalised "ground state" of the system elements, i.e. the state with minimum energy and complexity (corresponding, by convention, to $n = 0$):

$$H_0(\xi)\psi_0(\xi) + \sum_n V_{0n}(\xi)\psi_n(\xi) = \eta\psi_0(\xi), \qquad (9a)$$

$$H_n(\xi)\psi_n(\xi) + \sum_{n' \neq n} V_{nn'}(\xi)\psi_{n'}(\xi) = \eta_n \psi_n(\xi) - V_{n0}(\xi)\psi_0(\xi), \qquad (9b)$$

where now $n, n' \neq 0$ (also everywhere below) and $\eta \equiv \eta_0 = E - \varepsilon_0$.

It is interesting to note that exactly the same system of equations is obtained by a similar procedure for apparently much simpler system configuration, where one has just two distributed entities ("fields") interacting with each other [15,17,18,20-24]:

$$[h_g(\xi) + V(\xi, q) + h_e(q)] \Psi(\xi, q) = E \Psi(\xi, q). \qquad (10)$$

This existence equation describes, for example, the dynamics of our world emergence and behaviour at its most fundamental, "quantum" levels, in the process of attraction of two initially homogeneous "protofields" of different ("gravitational" and "electromagnetic") physical nature, described by the generalised Hamiltonians $h_g(q)$ and $h_e(q)$ respectively [15,16,20-24]. However, if we imagine that those distributed interacting entities have their internal structure (local inhomogeneities), then the system configuration may be not really different from the above "explicitly many-body" problem, which explains the coincidence of the transformed formulation of both problems in terms of the element degrees of freedom, eqs. (5) and (9). The unreduced brain dynamics can also be considered as a result of interaction between the distributed electromagnetic and chemical components, though provided with the developed super-structure within the neuron network [15,16]. This analogy between the lowest and highest levels of world dynamics has not only formal, but profound physical meaning, and we shall continue to reveal its manifestations below.

Expressing $\psi_n(\xi)$ from eqs. (9b) with the help of the standard Green function technique [25,26] and inserting the result into eq. (9a), we reformulate the problem in terms of *effective* existence equation formally involving only "common" ("structural") degrees of freedom ($\xi$) [15-21]:

$$[h_0(\xi) + V_{\text{eff}}(\xi;\eta)]\psi_0(\xi) = \eta\psi_0(\xi) , \qquad (11)$$

where the *effective (interaction) potential (EP)*, $V_{\text{eff}}(\xi;\eta)$, is given by

$$V_{\text{eff}}(\xi;\eta) = V_{00}(\xi) + \hat{V}(\xi;\eta), \ \hat{V}(\xi;\eta)\psi_0(\xi) = \int_{\Omega_\xi} d\xi' V(\xi,\xi';\eta)\psi_0(\xi'), \quad (12a)$$

$$V(\xi,\xi';\eta) \equiv \sum_{n,i} \frac{V_{0n}(\xi)\psi_{ni}^0(\xi)V_{n0}(\xi')\psi_{ni}^{0*}(\xi')}{\eta - \eta_{ni}^0 - \varepsilon_{n0}} , \ \varepsilon_{n0} \equiv \varepsilon_n - \varepsilon_0 , \quad (12b)$$

and $\{\psi_{ni}^0(\xi)\}$, $\{\eta_{ni}^0\}$ are the complete sets of eigenfunctions and eigenvalues of an auxiliary, truncated system of equations (recall that $n,n' \neq 0$):

$$H_n(\xi)\psi_n(\xi) + \sum_{n' \neq n} V_{nn'}(\xi)\psi_{n'}(\xi) = \eta_n\psi_n(\xi) . \qquad (13)$$

The general solution of the initial existence equation, eq. (2), is then obtained as [15,16,20,26]:

$$\Psi(\xi,Q) = \sum_i c_i \left[ \Phi_0(Q) + \sum_n \Phi_n(Q)\hat{g}_{ni}(\xi) \right] \psi_{0i}(\xi) , \qquad (14)$$

$$\psi_{ni}(\xi) = \hat{g}_{ni}(\xi)\psi_{0i}(\xi) \equiv \int_{\Omega_\xi} d\xi' g_{ni}(\xi,\xi')\psi_{0i}(\xi') ,$$

$$g_{ni}(\xi,\xi') \equiv V_{n0}(\xi')\sum_{i'} \frac{\psi_{ni'}^0(\xi)\psi_{ni'}^{0*}(\xi')}{\eta_i - \eta_{ni'}^0 - \varepsilon_{n0}} , \qquad (15)$$

where $\{\psi_{0i}(\xi)\}$ are the eigenfunctions and $\{\eta_i\}$ the eigenvalues found eventually from the effective dynamic equation, eq. (11), while the coefficients $c_i$ should be determined from the state-function matching conditions along the boundary where interaction vanishes. The observed system density, $\rho(\xi,Q)$, is given by the squared modulus of the state-function ampli-tude:[1] $\rho(\xi,Q) = |\Psi(\xi,Q)|^2$.

Reformulation of the initial problem, eqs. (1)-(10), in the "effective" version of eqs. (11)-(15) is known in scattering theory and related solid-state theory applications as the method of optical, or effective, potential (see e.g. [25]). The main "difficulty" of a nonintegrable problem is not really resolved in it, but rather displaced from the state-function equation as such to a more complicated potential in a simpler, *externally* integrable equation: the effective, or "optical", potential it contains depends on the unknown problem solutions and therefore bares the full problem difficulty. However, the "effective" problem formulation, being technically equivalent to the original one, has the advantage of much more detailed, explicitly expressed dynamical content (cf. e.g. eqs. (9) and (11)-(15)) revealing if not the desired solution, but at least its main dynamical components and their recurrent, interaction-driven entanglement. The conventional theory proceeds by very poor use of these advantages and prefers, according to its dominating paradigm, to cut severely the "nonintegrable" EP expressions in order to end up with a "closed", analytically finite, or "exact", solution within a version of "perturbation theory" (e.g. [25,27]). In such a theory the essential, properly dynamic content of the interaction process, reflected just in the details of EP expression, eqs. (12), is thrown off, and the system configuration is obtained as a rather straightforward, mechanistic reproduction of the remaining, reduced potential shape. Correspondingly, the result thus obtained cannot contain nontrivial dynamical effects, replaced by mere geometric, generally small "deformation" of the initial configuration and the simple mechanical sum, rather than true entanglement, of the system components (this reductive logic also gives rise to the popular concept of "geometrisation of physics"). And although all perturbative expansions appear to be mathematically incorrect for any real problem, while the studied sys-tems do show dynamically involved, non-mechanistic and non-geometric features (up to intelligence and consciousness), the reductive approximation of perturbation theory dominates in scholar science approach even to most complicated systems, probably just due to its simplicity and despite the glaring contradiction to the observed explicitly complex behaviour, already in the simplest physical systems.

---

[1] This rule corresponds to so-called "wave-like" (undulatory) levels of complex dynamics [15], where the main entities have a distributed and compressible physical structure and are described by wave equations using, in general, complex-number presentation. Those undulatory levels alternate with "particle-like", or "classical", levels of complexity, where the main entities have a permanently localised, "hard" structure and the measured quantities like "generalised density" are derived from the state-function amplitude itself, $\rho(\xi,Q) = \Psi(\xi,Q)$ (it obeys now classical equations for real-valued, directly measurable distribution function). In the case of (truly) intelligent and conscious behaviour one deals with undulatory, "quantum" kind of behaviour at the main underlying levels of dynamics (see below for more details) and therefore one should use rather the "wave-like" relation between the state-function and measured "density", but these technical details do not influence the main conclusions about the fundamental origin and structure of complex system dynamics.

To fully realise the EP problem formulation and obtain its unreduced solution, we note the *self-consistent nonlinear* dependence of the EP equations, eqs. (11)-(12), on the eigen-solutions to be found, appearing *dynamically*, even for the formally linear initial dynamic equations (eqs. (1), (2), (5), (9), (10)). It is not difficult to show [15,16,20,21] that this *dynamic, essential nonlinearity* of a problem, remaining hidden in any its ordinary, straightforward formulation, gives rise to the qualitatively new phenomenon of *dynamic multivaluedness*, i.e. the *redundant number* of locally complete solutions which, being *equally real* and *mutually incompatible*, are forced, by the main system interaction itself, to permanently replace each other in the *causally random* order thus defined. Indeed, if we designate by $N_\xi$ and $N_Q$ the numbers of terms in the sums over *i* and *n* in eq. (12b) (equal to the number of system components *N* and the number of their internal states, respectively), then it follows that the total number of the problem eigen-solutions, determined by the maximum eigenvalue power in the characteristic equation, is $N_{\max} = N_\xi(N_\xi N_Q + 1) = (N_\xi)^2 N_Q + N_\xi$, which gives the $N_\xi$-fold redundance with respect to the "normal" complete set of $N_\xi N_Q$ eigen-solutions for the initial system of equations, eqs. (5), (9), and an additional, "incomplete" set of $N_\xi$ eigen-solutions. Other estimates of the number of solutions, using geometric, model, and simple "physical" considerations [15,24,26], show that the found additional solutions are all equally real (not spurious) and have generally similar origin and structure.

This conclusion is confirmed by observation of the ensuing chaotic change of states in various many-body systems and interaction processes, without a really consistent explanation for it within the standard, dynamically single-valued theory.[2] Therefore, based on the rigorously obtained

---

[2] The dynamically single-valued, or *unitary*, models used, in particular, in scholar versions of the "science of complexity" try to *imitate* system realisation multitude by various artificial constructions, such as "attractors", in *abstract*, mathematical "spaces", but those illusive structures are always "produced" by the *single* available system state and trajectory, i.e. without any real change of system configuration in the real space. As a result, various imitative structures of the

multivaluedness of unreduced many-body problem solution, we can state that its general, now *really complete* solution, can be presented, in terms of observable generalised density $\rho(\xi,Q)$, as the *causally probabilistic sum* of individual realisation densities, $\rho_r(\xi,Q)=|\Psi_r(\xi,Q)|^2$, numbered by index *r* here and below:

$$\rho(\xi,Q) = \sum_{r=1}^{N_\Re} {}^\oplus \rho_r(\xi,Q) , \qquad (16)$$

where $N_\Re$ ($= N_\xi = N$) is the total number of system realisations, and the sign $\oplus$ serves to designate the special, causally random character of summation. The nontrivial origin of the latter, which cannot have any correct analogy in the dynamically single-valued theory, involves the *unceasing*, explicit change of system configurations, occurring in the *truly random* (rigorously unpredictable and noncomputable) order and driven exclusively by the main, initially totally *deterministic* interaction between system components, the *same* one that shapes the details of each emerging realisation configuration (functions $\{\rho_r(\xi,Q)\}$).

Since we have discovered in that way the truly dynamic origin of (any) randomness inevitably generated in a real interaction process, we can also provide the related *purely dynamic definition of probability*, $\alpha_r$, of each *r*-th realisation emergence, obtained in the form:

$$\alpha_r = \frac{1}{N_\Re} \quad (r = 1,...,N_\Re) , \qquad \sum_{r=1}^{N_\Re} \alpha_r = 1 . \qquad (17a)$$

As in many practical cases those elementary system realisations are inhomogeneously grouped into larger, actually observed "super-realisations" (or compound realisations), the dynamic probability definition takes, in general, the following practically adapted form:

$$\alpha_r(N_r) = \frac{N_r}{N_\Re} \left( N_r = 1,...,N_\Re ; \sum_r N_r = N_\Re \right), \sum_r \alpha_r = 1 , \qquad (17b)$$

where $N_r$ is the number of "elementary" realisations in the *r*-th "compound" (actually observed) realisation.

---

unitary "science of complexity" represent at best only extremely limited, one- or zero-dimensional (point-like) projections of the real, dynamic multivaluedness [15,16]. This difference between the unreduced dynamic multivaluedness and its unitary imitations is especially important in such explicit complexity manifestations as intelligence and consciousness, whose very essence is given just by the detailed, fractally structured system configurations (see below) and their permanent change, rather than a smooth enough "trajectory" of a system with a fixed or "adiabatically" evolving configuration in the unitary theory.

It is evident that the obtained expression for realisation probabilities provides the *universal,* purely *dynamic* and rigorously derived (rather than postulated) *concept and definition of probability*, helplessly missing in the usual, empirically based, postulated probability notion (therefore the scholar statistical mechanics and related branches represent but a "probabilistic" aspect of the same, dynamically single-valued, unitary projection of reality). Correspondingly, the emergence and disappearance, or change, of successive realisations represents the rigorously specified and universal definition of *event* (see also below), another empirically postulated notion widely used in the unitary theory, but always escaping consistent specification. The qualitatively new, causally complete probability content thus derived, eqs. (17), is distinguished by the fact that it does not depend on the number of actually observed events or even any event observation at all: contrary to any conventional probability version, it remains valid even for a single (next) expected event or any their "statistically small" number. However, if the number of observed events does become statistically large, we can correctly define the *expectation (average) value* of the observed quantity:

$$\rho_{\exp}(\xi,Q) = \sum_{r=1}^{N_{\Re}} \alpha_r \rho_r(\xi,Q) \ . \tag{18}$$

A useful dynamic probability aspect is related also to the *generalised wavefunction*, introduced below.

The internal structure of realisation change process can be better seen if we rewrite in full detail the expressions for the unreduced EP and state-function, eqs. (12) and (14), for a given, *r*-th realisation:

$$V_{\mathrm{eff}}(\xi;\eta_i^r)\psi_{0i}^r(\xi) = V_{00}(\xi)\psi_{0i}^r(\xi) +$$

$$+ \sum_{n,i'} \frac{V_{0n}(\xi)\psi_{ni'}^0(\xi) \int_{\Omega_\xi} d\xi' \psi_{ni'}^{0*}(\xi') V_{n0}(\xi') \psi_{0i}^r(\xi')}{\eta_i^r - \eta_{ni'}^0 - \varepsilon_{n0}} \ , \tag{19}$$

$$\rho_r(\xi,Q) = |\Psi_r(\xi,Q)|^2 \ ,$$

$$\Psi_r(\xi,Q) = \sum_i c_i^r \left[ \Phi_0(Q) + \sum_n \Phi_n(Q) \hat{g}_{ni}^r(\xi) \right] \psi_{0i}^r(\xi) \ , \tag{20}$$

$$\psi_{ni}^{r}(\xi) = \hat{g}_{ni}^{r}(\xi)\psi_{0i}^{r}(\xi) = \int_{\Omega_{\xi}} d\xi' g_{ni}^{r}(\xi,\xi')\psi_{0i}^{r}(\xi'),$$

$$g_{ni}(\xi,\xi') = \sum_{i'} \frac{\psi_{ni'}^{0}(\xi)V_{n0}(\xi')\psi_{ni'}^{0*}(\xi')}{\eta_{i}^{r} - \eta_{ni'}^{0} - \varepsilon_{n0}}.$$

As can be seen from eqs. (19)-(20), the same resonant denominator structure that gives the unreduced EP multivaluedness, eq. (19), explains the structure of each realisation, eq. (20), that tends to concentrate around a particular location, given by the corresponding eigenvalue $\eta_{i}^{r}$ (it can be conveniently marked as $\eta_{r}^{r}$), due also to the "cutting" action of integrals in the numerator. The system in each particular realisation as if "digs" a dynamic potential pit for itself, where it temporarily falls, until the well and related system localisation disappear in favour of a transient delocalisation in a specific "intermediate" state called also the "main" realisation and common for all "regular", localised realisations, before falling into the next "regular", compact realisation with another, randomly chosen centre of localisation, and so on. This unceasing realisation change and related qualitative change of system configuration and properties, forming the universal basis for any real, *dynamically multivalued (chaotic) structure formation*, results from the intrinsic, irreducible, and *permanently* present *dynamic instability* of a real system interaction process, revealed explicitly by the unreduced EP formalism in the form of nonlinear feedback interaction loops (self-consistent EP dependence on the eigen-solutions to be found) and absent in any perturbative, "exact" solutions obtained just by cutting those essential links (they also remain "hidden" in any straightforward problem formulation, such as eqs. (1), (2), (5), (9), and (10)). The important relation of this totally "spontaneous" structure emergence to the fundamental, dynamic *origin of time* is considered below.

As for the mentioned specific, delocalised system realisation, it corresponds to the "incomplete" set of eigenvalues revealed above in the analysis of the total number of eigenvalues and can be explicitly obtained from the effective existence equation, eq. (11), as a particular solution for which, contrary to all other solutions, the EP magnitude is indeed close to its weak-interaction, separable value, $V_{\text{eff}}^{0}(\xi;\eta_{i}^{0}) \simeq V_{00}(\xi)$. Therefore the "main" realisation is the direct analogue, within the unreduced, dynamically multivalued description, of the single realisation remaining in the usual, dynamically single-valued theory, where it realises the averaged, "statistical" projection of the multivalued, permanently changing dynamics to the limited, zero-dimensional space of a unitary "model". The specific role of the intermediate realisation in the multivalued system dynamics outlined above corresponds to its properties of the *generalised wavefunction*, or *distribution function*, with its causally explained chaotic structure [15,16,20-23]. It takes the form of ordinary quantum-mechanical wavefunction at the lowest, quantum levels of world complexity (now causally understood without any esoteric "mysteries"), but is defined also for any other complexity level, where it can be closer to the quantum wavefunction properties for "wave-like" levels of complexity (e.g. brain dynamics, see below) or closer to the (extended) classical "distribution function" for "particle-like" complexity levels (with permanently localised interacting entities).

## 2.2. Dynamic entanglement and multivalued fractality

The described structure of realisation change process involves also the phenomenon of *dynamic entanglement* of interacting system components, which is inseparably related to the major feature of dynamic multivaluedness of the unreduced system dynamics and expressed formally by the dynamically involved products of functions of $\xi$ and $Q$ in the state-function expressions, eqs. (14), (20). Dynamic entanglement specifies the abstract property of "nonseparability" of the unitary theory: any real system is "nonseparable" just because the degrees of freedom of interacting components are physically, dynamically "entangled" ("woven") with each other into a permanently changing system configuration. Therefore the whole interaction process and its results can be described as *dynamically multivalued entanglement* of interacting entities, where component entanglement constitutes each "regular" realisation and during realisation change the components first transiently disentangle, forming the quasi-free state of "intermediate" realisation (generalised wavefunction, see the previous Section), and then entangle again in a new version of system configuration (another regular realisation). In that way one obtains the real, physically tangible and permanently internally changing "tissue of reality", constituting the "flesh" of any real system or structure, while in the unitary theory the latter is replaced by its abstract, illusive and "weightless" envelope of "separated variables", constituting the essence of all imitative, "exact" solutions and "integrable" models.

The dynamic entanglement and physical nonseparability of real system structure have also the important dimension of *dynamical, multivalued (probabilistic) fractal*. Indeed, the unreduced problem solution, eqs. (11)-(17), contains explicitly only one level of system splitting into incompatible and permanently changing realisations, while it refers also to unknown solutions of the "auxiliary" system of equations, eqs. (13). In principle, after having revealed the major, universally nonperturbative effect in the form of dynamic multivaluedness, we have some freedom to use an approximate solution for *this*, auxiliary system and obtain its eigen-solutions $\{\psi_{ni}^0(\xi), \eta_{ni}^0\}$ entering the main formulas (eqs. (12), (14), (19), (20)) from a reduced, "integrable" version of eqs. (13), such as

$$\left[H_n(\xi) + \tilde{V}_n(\xi)\right]\psi_n(\xi) = \eta_n \psi_n(\xi), \qquad (21a)$$

where the ordinary, single-valued potential may vary within some more or less evident borders:

$$\left|V_n(\xi)\right| \leq \left|\tilde{V}_n(\xi)\right| \leq \left|\sum_{n'} V_{nn'}(\xi)\right|. \qquad (21b)$$

In this case we limit our attention to the first, main level of multivalued dynamics and ignore its further involvement hidden in the unreduced solution of the auxiliary system of equations. If, however, we want to continue the study of the real, non-simplified system dynamics, we can avoid the above approximation and apply the same EP method of solution to eqs. (13).

Separating explicitly the equation for $\psi_n(\xi)$ in eqs. (13), we rewrite the auxiliary system in the form analogous to eqs. (9) for the main system:

$$H_n(\xi)\psi_n(\xi) + \sum_{n' \neq n} V_{nn'}(\xi)\psi_{n'}(\xi) = \eta_n \psi_n(\xi), \qquad (22a)$$

$$H_{n'}(\xi)\psi_{n'}(\xi) + \sum_{n'' \neq n, n'} V_{n'n''}(\xi)\psi_{n''}(\xi) = \eta_n \psi_{n'}(\xi) - V_{n'n}(\xi)\psi_n(\xi), \ n' \neq n. \qquad (22b)$$

Expressing now $\psi_{n'}(\xi)$ through $\psi_n(\xi)$ from eqs. (22b) with the help of the Green function for its truncated, "homogeneous" part and inserting the result into eq. (22a), we arrive at the "effective" formulation for the auxiliary system of equations taking now an "integrable" configuration similar to that of eq. (11):

$$\left[ h_0(\xi) + V_{\text{eff}}^n(\xi;\eta) \right] \psi_n(\xi) = \eta_n \psi_n(\xi), \tag{23}$$

where

$$V_{\text{eff}}^n(\xi;\eta) = V_{nn}(\xi) + \hat{V}_n(\xi;\eta), \quad \hat{V}_n(\xi;\eta)\psi_n(\xi) = \int_{\Omega_\xi} d\xi' V_n(\xi,\xi';\eta)\psi_n(\xi'), \tag{24a}$$

$$V_n(\xi,\xi';\eta) = \sum_{i,n'\neq n} \frac{V_{nn'}(\xi)\psi_{n'i}^{0n}(\xi)V_{n'n}(\xi')\psi_{n'i}^{0n*}(\xi')}{\eta_n - \eta_{n'i}^{0n} + \varepsilon_{n0} - \varepsilon_{n'0}}, \tag{24b}$$

and $\{\psi_{n'i}^{0n}(\xi), \eta_{n'i}^{0n}\}$ are the eigen-solutions of a yet more truncated auxiliary system of the next level:

$$H_{n'}(\xi)\psi_{n'}(\xi) + \sum_{n''\neq n} V_{n'n''}(\xi)\psi_{n''}(\xi) = \eta_{n'}\psi_{n'}(\xi), \quad n' \neq n. \tag{25}$$

We can obviously continue this process further, obtaining each time ever more truncated system of auxiliary equations, until we remain with only one equation for a single mode, which is solved explicitly and terminates the real process of dynamical fractal formation.

It is important that at each level of fractal hierarchy we have the same phenomenon of dynamically multivalued entanglement generated by the same dynamically nonlinear feedback mechanism as the one revealed above for the main level of splitting and described now by the unreduced EP formalism of eqs. (23)-(24). This means that, contrary to the conventional, dynamically single-valued fractals (including their artificially "stochastic" versions), each level of the unreduced fractal hierarchy contains *permanent change* of realisations in a *dynamically* random order [15-17]. As a result, such real fractal becomes a permanently, *coherently* moving and *adaptively* developing, "living" arborescent structure representing the really complete solution of the many-body problem in its full complexity. It can be expressed as a "multi-level" causally probabilistic sum (cf. eq. (16)):

$$\rho(\xi,Q) = \sum_{j=1}^{N_f} \sum_{r=1}^{N_{\Re j}} {}^\oplus \rho_{jr}(\xi,Q), \tag{26}$$

where $\rho_{jr}(\xi,Q)$ is the measured quantity for the $r$-th realisation at the $j$-th level of dynamic fractality, $N_{\Re j}$ is the number of (observable) realisations at the $j$-th level, and $N_f$ is the final or desired level number. This expression is accompanied by the corresponding dynamic definitions of probabil-ity and expectation values for each level of fractal hierarchy, analogous to eqs. (17), (18), which we do not reproduce here.

The dynamic entanglement at each level of fractality endows the unreduced fractal with "flesh and blood" specific for the given system and determining the perceived detailed *"quality"*, or *texture*, of system structure. The latter is directly related to the problem *nonseparability*, acquiring now a transparent *physical* meaning (it is impossible to separate *fractally* entangled components forming *permanently changing, unstable* realisations) and actually underlying the *real* system *existence* itself.[3] The unreduced dynamical fractal can be described by a number of slightly different versions of the same EP method (depending on the chosen "zero-th approximation" etc.), but they all give the same fundamental result, describing the unreduced system behaviour as a dynamically probabilistic hierarchy of permanently changing and internally entangled realisations. It is clear that due to the hierarchy of levels of dynamical splitting the total number of system realisations is exponentially large (where already the argument of the exponential function will be a large number for any real multi-component system), which determines the huge dynamic efficiency of the unreduced dynamic fractality playing the key role in various applications, including intelligence dynamics (Section 3.2). On the other hand, and this is another side of "living" structure efficiency, the probabilistic dynamical fractal always preserves its *integrity (wholeness)* and forms and changes as an *intrinsically unified* configuration of the entire interaction process.

## 2.3. Unified classification of dynamic regimes: From global chaos to multivalued self-organisation

Since the existing world structures at any scale result from the corresponding interaction process development, it is clear that the entire universe, or any its part, can be considered as the single, dynamically unified, probabilistic fractal structure (see the previous Section), where the emerging more "solid" (distinct) branches (at a certain level) correspond to "interacting objects", whereas the finely structured fractal "foliage" around them constitutes the well-specified, *material* content of "interaction (potential)" as such. Therefore, contrary to the simplified symmetries of usual fractals (scale invariance) and their limited number of prototype real objects, the unreduced dynamical fractal represents the *exact* structure and dynamics of *any* kind of object and can show approximate scale invariance, or any other particular kind of structure, only within a limited range of scales. However, the huge diversity of possible dynamic regimes can now be classified as the combination of two limiting cases, designated as *uniform, or global, chaos* and *(dynamically multivalued) self-organisation, or self-organised criticality (SOC)*.

---

[3] It shows, in particular, that all basically separable, "exact" solutions and dynamically single-valued models and concepts of the unitary science can *never* describe the real system as it is, in its *essential*, major *quality*, providing instead just a zero-dimensional, point-like version of an external, "immaterial", abstract system shape.

To demonstrate the origin of both regimes, we note that in the limit of small eigenvalue separation (frequency) for the chosen structure-dependent, or "external", degrees of freedom ($\xi$) with respect to those for the element-dependent (internal) degrees of freedom ($Q$), $\Delta\eta_i \ll \Delta\eta_n \sim \Delta\varepsilon$, or $\omega_\xi \ll \omega_Q$ (where $\Delta\eta_i$ and $\Delta\eta_n$ are the eigenvalue separations with respect to changing *i* and *n* respectively in eq. (12b), and $\omega_\xi$ and $\omega_Q$ are the corresponding frequencies), the summation over *i* in the general EP expression, eq. (12b), can be performed independently in the numerator, giving a local and single-valued EP limit (in view of the completeness of the auxiliary equation solution set) [16]:

$$V(\xi,\xi';\eta) = \delta(\xi-\xi') \sum_n \frac{|V_{0n}(\xi)|^2}{\eta - \eta_{ni}^0 - \varepsilon_{n0}},$$

$$V_{\text{eff}}(\xi;\eta) = V_{00}(\xi) + \sum_n \frac{|V_{0n}(\xi)|^2}{\eta - \eta_{ni}^0 - \varepsilon_{n0}}.$$
(27)

Similar results are obtained for the state-function, starting from eqs. (14)-(15) [16]. This is the limiting regime of self-organisation, giving a distinct and "regular" system structure. However, it always remains only an approximation to reality, and the unreduced EP deviations from the limit of eqs. (27), however small they are, have a *qualitatively strong* character: the real EP and state-function are composed from *many* close (similar) enough and very quickly changing, but nevertheless different, realisations, which means that *any* real self-organisation, and the resulting "distinct" structure, has *dynamically multivalued*, internally *chaotic*, *fractal* character and *permanently* (and randomly) fluctuates, in a large range of scales, around the observed "average" shape, thus *comprising* and *extending* the phenomenon of *self-organised criticality (SOC)*, which otherwise suffers, in its standard version, from a conflict with intrinsic chaoticity and separation from other cases of "self-organisation" [15-19]. The internal chaotic realisation change within an externally "regular" structure constitutes, despite its "hidden" character, the true basis of that structure emergence and existence, without which it loses any realistic meaning (including its *proper time flow*, which is an ever persisting difficulty of the unitary theory). This limiting case unifies also the extended versions of all other unitary imitations of dynamically multivalued SOC, such as "control of chaos", "synchronisation", "phase locking", etc., remaining split and incomplete in their usual versions.

The opposite limiting case of *uniform, or global, chaos*, is realised when the above characteristic system frequencies (or eigenvalue separations) are close to each other, $\Delta\eta_i \simeq \Delta\eta_n \sim \Delta\varepsilon$, or $\omega_\xi \simeq \omega_Q$, i.e. the corresponding degrees of freedom fall in resonance. In that case the individual realisations eigen-solutions are so entangled among them that there is no possibility to separate them, even approximately, and the permanent, chaotic realisation change takes its explicit, externally visible form, where *sufficiently different* realisations change at a not too fast and not too slow rate close to the main system frequencies. One obtains thus the *universally* applicable *criterion of global, explicit chaoticity* that coincides with the *condition of resonance* between the main system motions [15,16,19]:

$$\kappa \equiv \frac{\Delta\eta_i}{\Delta\eta_n} = \frac{\omega_\xi}{\omega_q} \simeq 1 \ , \qquad (28a)$$

where the parameter of chaoticity, $\kappa$, is introduced by this definition. Note that in that way we also clarify the true meaning of the "familiar" phenomenon of resonance itself, inevitably omitted in its conventional, perturbative description. Correspondingly, the condition

$$\kappa \ll 1 \qquad (28b)$$

(as well as $\kappa \gg 1$) provides the universal criterion of occurrence of (multivalued) SOC kind of dynamics and global "regularity", i.e. absence of pronounced, externally dominating chaoticity. Note that at $\kappa \gg 1$ one obtains just another kind of chaotic mode enslavement within an externally regular shape (or multivalued SOC), which is "complementary" with respect to that obtained at $\kappa \ll 1$ and usually only one of them represents a major interest within each particular problem.

*Universality* of the criterion of eqs. (28) is of particular interest for the unreduced science of complexity, since it provides a simple and unified *principle of classification* of *all* possible kinds of behaviour and dynamics of *any* system, constituting a confusing problem for the scholar theory. It implies also that system behaviour can gradually vary between those too extreme cases of "global regularity" and "global chaoticity", depending on the value of chaoticity parameter $\kappa$. These statements are confirmed by independent analysis of the particular case of true quantum chaos (and its correct classical limit) [24], where the corresponding parameter of transition to global chaos, $K$, is directly related to $\kappa$, $K = \kappa^2$. The conceptual and technical transparency of the proposed criterion of chaoticity and regularity is to be compared with obscurity of its unitary imitations, containing incorrect statements and technical trickery. In that way the genuine, intrinsic complexity of unreduced, multivalued dynamics underlies the universal simplicity of the key criteria formulation and related harmony of the general picture, whereas the illusive simplicity of dynamically single-valued, perturbative "models" leads inevitably to technical and conceptual uncertainty, leaving no hope for universally applicable, realistic understanding. In particular, the above two limiting regimes of unreduced complex dynamics, as well as their universal meaning and relation, appear to be indispensable for understanding of the emerging phenomena of intelligence and consciousness and their internal dynamics (see Sections 3 and 4).

## 2.4. Universal definition, symmetry and formalism of unreduced dynamic complexity

The rigorously expressed *notion and quantity of dynamic complexity* as such can be *universally* defined now in terms of the above unreduced interaction analysis as any growing function of the total (or observable) number of system realisations, or related rate of their change, equal to zero for (actually unrealistic) case of only one realisation [15-19]:

$$C = f(N_\Re), \quad \frac{df}{dN_\Re} > 0, \quad f(1) = 0 , \qquad (29)$$

where $C$ is a quantitative measure of complexity and $f(x)$ is an arbitrary function with the designated properties. An "integral" measure of complexity is provided by the popular "logarithmic" expression, $C = C_0 \ln(N_\Re)$, which properly reflects the hierarchical structure of complexity, but acquires its true meaning and usefulness only in combination with the universally nonperturbative analysis of the underlying interaction process that specifies clearly the relevant system realisations. Various "differential" measures of complexity are provided by rates of unceasing realisation change (temporal or spatial), taking the form of familiar quantities, such as mass, energy, or momentum, but now provided with a quite new, causally complete and universal meaning (see below). It is evident from the above picture that the unreduced, dynamically multivalued complexity basis thus defined includes also the notion of *chaoticity*, although the actually observed, *apparent* degree of *irregularity* for a particular case may vary depending on the specific regime of complex dynamics (but the unreduced, internal chaoticity is *always* there and proportional to the unreduced complexity, see also the generalised entropy definition below).

Note that, according to the definition of eq. (29), the unreduced, "genuine" dynamic complexity of any dynamically single-valued "model" from usual theory (including all its versions of the "science of complexity") is strictly zero, the single available realisation of this unitary projection being presented usually by an "averaged" structure that corresponds to the "main" realisation (or "generalised wavefunction") of the unreduced picture. However, that zero-dimensional projection of the unitary theory may have a structure, and various really observed or arbitrarily postulated, often purely abstract elements of that point-like structure are often substituted for real system realisations, existing in real space, after which a non-zero (but totally false) value of complexity is readily obtained by formal application of the same expressions (e.g. the "logarithmic" complexity measure). In addition, the notion of complexity is inevitably confused, within the unitary framework, with various "similar" notions, such as "information", "entropy" and "chaoticity" (see ref. [15,16] and below for more details).

An inquiry into the detailed structure of complexity brings us to the *dynamic origin of space and time* hierarchy revealed by the above unreduced problem solution. Indeed, the totally "spontaneous" (autonomous) structure emergence, in the form of dynamic system "concentration" around each of its permanently changing realisations, *consistently derived* by the unreduced EP formalism, should then be considered as *real, physical* space structure emergence at the corresponding "level of complexity". The generalised "space point" of each complexity level is provided by the emerging realisation structure at the moment of its maximum dynamical squeeze (before system transition to the next realisation), given by eqs. (19)-(20), with the centre of this "point-structure" being designated by the corresponding eigenvalue, $\eta_r^r$ (see the discussion after eqs. (19), (20)). The characteristic size, $r_0$, of this real space element is given by the eigenvalue separation, $\Delta\eta_n$, with respect to the "internal" degrees of freedom ($Q$), within one realisation: $r_0 \simeq \Delta\eta_n$ (it is assumed here that $\Delta\eta_n$ is measured in the same units as the corresponding "wavelength"). A yet more important space dimension, the elementary distance (length element, or characteristic wavelength), $\Delta x = \lambda$, emerges *dynamically* in the form of eigenvalue separation with respect to "external" degrees of freedom ($\xi$) or neighbouring realisations: $\Delta x \simeq \Delta\eta_i \simeq \Delta\eta_r^r$. This is the spatial measure, or "size", of a single system jump between its successive realisations. The dynamically emerging *time element* measures the "intensity", actually given by *frequency*, $\nu$, of realisation change, which is inversely proportional to the direct time "distance" (or *period*), $\Delta t = \tau = 1/\nu$, between two successive *events* of realisation emergence specified by the unreduced EP formalism, which can be independently estimated as $\Delta t = \Delta x / c$ (where $c$ is the speed of material signal propagation in the initial, "structureless" system). In other words, the time element provides the dynamically emerging duration of system jump between two successive realisations.

Note that the space structure thus derived is intrinsically *discrete* (eventually due to the *wholeness* of unreduced interaction dynamics [15,16]), while time is fundamentally *irreversible* (because of the *dynamic unpredictability* of each next realisation) and *unceasingly flowing* (due to the same *dynamic multivaluedness*, driven by the main interaction process itself and thus unstoppable, if the system maintains its existence as such). Due to the dynamically fractal structure of any particular system and hierarchy of complexity in the whole, the real space and time have the corresponding hierarchic, fractal structure, with the proper dynamical links between successive levels (branches) of the dynamical fractal. The lowest, most fundamental level of space and time is provided by the interaction between two primordial, initially homogeneous, and physically real proto-fields that gives rise to (dynamically emerging) elementary particles and their interactions. Here $r_0$ is equal to the intrinsic particle size, such as the "classical radius of the electron", $\Delta x = \lambda = \lambda_\mathrm{C}$ is the Compton wavelength, and $\Delta t = \tau = h/m_0 c^2$ is the internal "quantum beat" period of the particle [15,16,20-23] (where $h$ is Planck's constant, $m_0$ is the particle rest mass, and $c$ is the speed of light). Whereas the space and time elements at each level are *dynamically* related among them, they are also qualitatively different from each other by their origin and role: space determines the *tangible*, "material" system structure, texture, or *specific "quality"* (including the *dynamically entangled* structure of each regular realisation forming the space element), while time has an *immaterial* nature (contrary to its incorrect "mixture" with space in the unitary science framework) and characterises the *intensity* of unceasing, irreducible *change* of that material space structure.

It follows that space and time thus universally and dynamically defined by the unreduced interaction process constitute two major, universal *forms of complexity* that can take a variety of different shapes in particular systems and at various levels of complexity. Space and time are directly made by the successively emerging and changing realisations of any real system, and therefore one can say that these two basic forms of complexity and their dynamic relation determine everything in the existing world structure. By contrast, various *measures of complexity* introduced above (starting from eqs. (29)) are suitable *functions* of realisation number or rate of change and thus of space and time, which provides the fundamental, *dynamically specified* origin of the very *notion of function*, usually considered in its abstract, mathematical meaning. Since the simplest possible combination of space and time, independently proportional to both space and time, is given by *action*, we arrive at the *extended interpretation of action* as a universal, integral *measure of unreduced dynamic complexity*, thus incorporating its *essentially nonlinear* origin and entangled internal structure:

$$\Delta \mathcal{A} = -E \Delta t + p \Delta x \; , \qquad (30)$$

where $p$ and $-E$ are initially just coefficients relating the dynamically determined increments of space $\Delta x$ and time $\Delta t$ to the increment of action $\Delta \mathcal{A}$. The analogy to the well-known relations from classical mechanics (where our *universal* description should remain valid) immediately shows, however, that $p$ and $E$ can be identified with the system *momentum* and *(total) energy*, respectively, now in their universally extended versions of *differential measures of complexity*:

$$E = -\frac{\Delta \mathcal{A}}{\Delta t}\bigg|_{x=\text{const}} \simeq \frac{\mathcal{A}_0}{\tau} \; , \qquad (31)$$

$$p = \frac{\Delta \mathcal{A}}{\Delta x}\bigg|_{t=\text{const}} \simeq \frac{\mathcal{A}_0}{\lambda} \; , \qquad (32)$$

where $\mathcal{A}_0$ is the magnitude of the characteristic increment (and value) of action for the given system and level of complexity.

The discrete increment of action-complexity (equal to Planck's constant with the negative sign, -*h*, at the lowest, quantum complexity level [15,16,20-23]) describes an elementary, indivisible step of system complexity "development" as its structure emerges in the driving interaction process. Appearing structural elements start interacting among them through the fractal net of interaction links, giving rise to higher-order and eventually higher-level structures. Every real change in this hierarchy of creation corresponds to a *negative* increment, or *decrease*, of action-complexity, or *dynamic information* ($\Delta \mathcal{A} < 0$), whereas another universal measure of complexity, *generalised dynamical entropy S*, simultaneously *increases* by the amount lost by action, so that their sum, the *total system complexity C*, remains constant during (closed) system evolution [15-19],

$$C = \mathcal{A} + S = \text{const}, \qquad (33a)$$

$$\Delta S = -\Delta \mathcal{A} > 0 \; . \qquad (33b)$$

This *universal law of conservation, or symmetry, of complexity*, determining evolution and existence of *any* system, from elementary particle to the universe and conscious brain, has a transparent physical meaning, where action-complexity describes available stock of "potential", latent form of initial interaction complexity (generalised, integral version of "potential energy") that transforms, by system evolution during interaction development, to the explicit, final form of fully developed system structure and dynamics represented by complexity-entropy (generalised, integral version of "kinetic" and "heat" energy). Entropy, as a measure of chaoticity, can only grow because of the fundamental dynamic uncertainty at every single step revealed above, but this is possible only at the expense of equally decreasing action-complexity that provides the universal "driving force" for the dynamic structure (entropy) creation. Because of such role of action-complexity, it is also called dynamic information and provides thus the correct, complex-dynamic extension of the notion of information (confused with entropy in usual theory). In that way, the universal science of complexity considerably extends and puts in order various reduced, often erroneous ideas of unitary science about complexity, entropy, information and relations between them [15,16].

Now, in order to find the universal dynamic expression of the symmetry of complexity, we can divide the differential form of the complexity conservation law, eq. (33b), by $\Delta t|_{x=\text{const}}$ to obtain the *generalised Hamilton-Jacobi equation* [15,16,18]:

$$\frac{\Delta \mathcal{A}}{\Delta t}\Big|_{x=\text{const}} + H\left(x, \frac{\Delta \mathcal{A}}{\Delta x}\Big|_{t=\text{const}}, t\right) = 0 , \qquad (34a)$$

where the Hamiltonian, $H = H(x,p,t)$, expresses a differential measure of the explicit, entropian complexity form, $H = (\Delta S/\Delta t)|_{x=\text{const}}$, and one deals with the *dynamically discrete* versions of partial derivatives giving energy and momentum, eqs. (31), (32). Expanding the Hamiltonian dependence on momentum in a power series,

$$H(x,p,t) = \sum_{n=0}^{\infty} h_n(x,t) p^n ,$$

where the expansion coefficients, $h_n(x,t)$, can be, in principle, arbitrary functions, we obtain the universal Hamilton-Jacobi equation in the form

$$\frac{\Delta \mathcal{A}}{\Delta t}\Big|_{x=\text{const}} + \sum_{n=0}^{\infty} h_n(x,t) \left(\frac{\Delta \mathcal{A}}{\Delta x}\Big|_{t=\text{const}}\right)^n = 0 , \qquad (34b)$$

where its coincidence with many particular equations for various $h_n(x,t)$ and series truncations becomes evident, especially if we rewrite it in terms of usual, continuous-limit symbols for partial derivatives:

$$\frac{\partial \mathcal{A}}{\partial t} + \sum_n h_n(x,t) \left(\frac{\partial \mathcal{A}}{\partial x}\right)^n = 0 . \qquad (34c)$$

Note that functions $h_n(x,t)$ here can have an additional dependence on $\mathcal{A}$, either through "potential energy" in the Hamiltonian or due to the eventual EP dependence on the solutions to be found in the actually implied, effective form of the formalism (see eqs. (12)-(15), (19)-(20)).

The unreduced, dynamically multivalued system evolution contains also phases of transition between realisations through the extended state of "generalised wavefunction" (or intermediate realisation), where the above expression in terms of action, reflecting the regular, "condensed" realisation quality, becomes inexact. The wavefunction state can be properly taken into account if we note that transitions between regular and intermediate realisations can also be considered as system structure development by transitions between neighbouring complexity sublevels, where the total complexity *C*, expressed by the product of complexity-entropy (regular realisations) and *wavefunction* $\Psi$ itself, should remain constant, $C = S\Psi = \text{const}$, meaning also that $\mathcal{A}\Psi = -S\Psi = \text{const}$. Therefore $\Delta(\mathcal{A}\Psi) = 0$ during one cycle of realisation change, which expresses the physically transparent condition of structural permanence of the *unique* intermediate realisation and leads to the following *universal* and *dynamically derived (causal) quantisation rule* [15,16,18]:

$$\Delta \mathcal{A} = -\mathcal{A}_0 \frac{\Delta \Psi}{\Psi} , \qquad (35)$$

where $\mathcal{A}_0$ is a characteristic action value that may also contain a numerical constant reflecting specific features of a given complexity level. We see that the relation between action and wavefunction, which takes the form of standard (Dirac) quantisation rules at the lowest (quantum) levels of complexity, can now be causally explained (contrary to "mysterious" postulates in the standard quantum theory) as expression of (physically real) realisation change dynamics and thus extended to any complexity level.

Substituting the obtained action expression through the wavefunction, eq. (35), into the generalised Hamilton-Jacobi equation, we get the respective forms of *generalised Schrödinger equation* [15,16,18]:

$$\mathcal{A}_0 \frac{\Delta \Psi}{\Delta t}\bigg|_{x=\text{const}} = \hat{H}\left(x, \frac{\Delta}{\Delta x}\bigg|_{t=\text{const}}, t\right)\Psi(x,t) , \qquad (36a)$$

$$\mathcal{A}_0 \frac{\Delta \Psi}{\Delta t}\bigg|_{x=\text{const}} = \sum_{n=0}^{\infty} h_n(x,t)\left(\frac{\Delta}{\Delta x}\bigg|_{t=\text{const}}\right)^n \Psi(x,t) , \qquad (36b)$$

$$\mathcal{A}_0 \frac{\partial \Psi}{\partial t} = \sum_{n=0}^{\infty} h_n(x,t) \frac{\partial^n \Psi}{\partial x^n} , \qquad (36c)$$

where the operator form of Hamiltonian, $\hat{H}$, is obtained from its functional form of eq. (34a) with help of the causal quantisation rule of eq. (35). If the Hamiltonian does not depend explicitly on time, we obtain the time-independent form of the universal Schrödinger equation:

$$\hat{H}\left(x, \frac{\Delta}{\Delta x}\bigg|_{t=\text{const}}\right)\Psi(x) = E\Psi(x) \,, \tag{36d}$$

where $E$ is the (constant) energy value. Note that the generalised Schrödinger formalism thus causally derived within the unreduced interaction process analysis is especially useful in description of unreduced intelligence and consciousness dynamics (see Section 3).

Another manifestation of the direct dynamical link between the common, delocalised state of wavefunction and different regular, "localised" system realisations takes the form of *generalised Born's probability rule*, which expresses a regular realisation probability, dynamically defined according to eqs. (17), through the wavefunction value for the corresponding system location (configuration) [15,16]. Similar to the above causal quantisation rule, the probability rule has a transparent physical meaning in the multivalued dynamics picture, since it states simply that the probability of wavefunction "reduction" (dynamical squeeze) to a particular realisation is proportional to the wavefunction magnitude around that particular realisation (and vice versa). In view of the permanent probabilistic transformation between the wavefunction and regular realisation, one could not imagine any other situation. One can derive the probability rule in a mathematically rigorous way by invoking the state-function matching conditions that should be used for evaluation of the coefficients $c_i^r$ in the general solution expression of eqs. (14)-(15) or (20) (see text after eqs. (15)). The state of wavefunction represents just that "dynamical border" of "quasi-free" system configuration, where the effective interaction is transiently "disabled" and the system "automatically" matches "itself to itself", but in a different state, i.e. it follows a "dynamic reconstruction" procedure (always driven by the same, major interaction). Therefore matching the state-function of eq. (20) in its "wavefunctional" phase to the corresponding "reduced" phase of a regular realisation (averaged over the internal degrees of freedom, unimportant here), we can see that the *r*-th realisation probability, $\alpha_r = \alpha(x_r) = \alpha(x)$, is given by both squared modulus of $c_i^r$ (properly averaged over $i$) and squared modulus of the wavefunction $\Psi(x)$:

$$\alpha_r = \alpha(x_r) = \alpha(x) = |\Psi(x)|^2 \,. \tag{37}$$

Note that in this form the probability rule is directly applicable to the "wave-like" levels of complexity (such as those of quantum behaviour and "subconscious" brain dynamics), whereas for levels with the dominating particle-like ("generalised classical") behaviour type one should use the generalised wavefunction, or distribution function, itself instead of its squared modulus.

The universal Hamilton-Jacobi and Schrödinger equations dynamically related by the causal quantisation condition and generalised probability rule constitute together the causally complete, *universal Hamilton-Schrödinger formalism*, eqs. (34)-(37), generalising *all* (correct) dynamic equations for particular systems [15,16,18]. The *unrestricted universality* of our description is *indispensable* for understanding of brain (intelligence) dynamics, since the latter obviously "reproduces" and thus encompasses *any* behaviour it can practically apprehend. Note that the explicitly "nonlinear" (in the usual sense) forms of the generalised Hamilton-Jacobi and Schrödinger equations, where functions $h_n(x,t)$ contain various (small) powers of action or wave function to be found, are often postulated in particular applications, but they are rather approximations to respective effective versions of initially "linear" equations, where such *essential, dynamic* nonlinearity appears, as we have seen, as a result of natural interaction loop development (see eqs. (12)-(15), (19)-(20) and the related discussion). Indeed, it is important that the above generalised equations include implicitly their unreduced, dynamically multivalued analysis and solution within the generalised EP method, constituting an essential extension with respect to usual, dynamically single-valued interpretation and solutions. Now we shall analyse manifestations of this universally defined complex behaviour and applications of the above description at the level of brain dynamics, including the emerging phenomena of intelligence and consciousness.

# 3. Intelligence and consciousness as unreduced complexity levels emerging in large and deep enough systems

## 3.1. Complex brain dynamics, generalised quantum beat and the brainfunction formalism

Note once again that the unrestricted *universality* of the above complexity derivation and concept, applicable to *both* real world dynamics *and* its reflection in an "intelligent" system of interacting elements ("generalised neurons"), plays a quite special, indispensable role in the ensuing theory of intelligence and consciousness, since that exact enough (and apparently unlimited) reflection of real world structure and dynamics is just the main distinctive feature of intelligent system behaviour. The latter can now be formally classified with the help of "complexity correspondence principle" [15,16], following in its turn from the universal symmetry of complexity (Section 2.4). This rule provides a rigorously specified expression of a rather evident fact that the full, unrestricted reproduction of a real (complex) behaviour pattern needs at least as much (or in practice even slightly more) complexity of the reproducing system dynamics. Despite its apparent simplicity, this rule has nontrivial practical applications and immediately shows, for example, that *all* "directly quantum" theories of brain function, appearing so readily in recent years and trying to explain it by the dynamics of the lowest, quantum levels of complexity (e.g. [10-12,29-39]), are *fundamentally deficient* and therefore *wrong*, irrespective of details, as well as *any unitary*, dynamically single-valued model of consciousness in terms of any system or level of world dynamics (such as many recent "physical" models of brain operation [39-46]). Indeed, in all those cases the (zero) level of unreduced complexity of the supposed (unitary) origin of consciousness is far below that of not only conscious, but even *any* real, *multivalued* system dynamics.

Returning to the unreduced interaction process that is at the origin of emerging, universally defined complexity (Section 2), we can now specify that interaction and its results for the case of natural or artificial brain (neural network) dynamics. We define here the generalised brain (intelligence) system as a system with a large enough number of *effectively* rather simple, in principle, interacting elements (each of them should typically have at least a few stable enough internal states), which are *massively* connected among them (details are to be specified below), thus realising their strong enough interaction that embraces *the entire system*. Our general "existence equation" for a system with unreduced interaction, eqs. (1) and (2), includes this case, but it can be further specified for the brain system in the following way taking into account explicit dependence on time (mainly due to interaction with the controlled environment):

$$\frac{\partial \Psi}{\partial t} = \left\{ h_0(\xi) + \sum_{k=1}^{N} \left[ h_k(q_k) + V_{0k}(\xi, q_k) \right] + \sum_{k=1, l>k}^{N} V_{kl}(q_k, q_l) \right\} \Psi(\xi, Q),$$

(38)

where the time variable *t* is suitably added to the independent variables (*Q*), so that the EP analysis remains practically unchanged (with the proper definition of generalised energies, e.g. in eqs. (3), (6)), including the basic system of equations (5), (9).

Note that eq. (38) generalises various model equations describing neural network dynamics (e.g. [47-49]), but due to its unrestricted universality it implies actually much more than neuron interaction through their direct, mechanical connection to each other. It involves the most fundamental, and quite indispensable, level of *global electro-chemical interaction* within and between natural brain neurons that should also have its analogue in any efficient system of genuine artificial intelligence and should be distinguished from the mere electromagnetic (e/m) interaction transmitted through connections between localised neurons. This latter interaction always exists in the brain, but it is essentially assisted there by interaction transmission through the biochemical cell connections and system-wide interaction between the two connection interfaces, the e/m and chemical ones. Recalling the analogy between the driving interaction processes in the brain and at the very first, quantum level of complex-dynamic structure emergence (see eq. (10)), we conclude that the unreduced brain dynamics is determined by the global, brain-wide, but highly inhomogeneous interaction between the e/m and chemical (physically real) "manifolds" constituted by all neurons and their connections, which is further assisted by individual inter-neuron couplings through both e/m and chemical cell connections [15,16]. Emergence of elementary particles and their interactions at much lower, quantum complexity levels are similarly described by interac-tion between the omnipresent e/m and gravitational protofields [15,16,20-23] (with the evident analogy between more "inert" behaviour of chemical and gravitational components of respective systems), but at that case the initial system configuration is effectively homogeneous, contrary to the very rugged "landscape" of the initial brain configuration.

It is this general analogy between the driving interaction configurations, as well as universality of the ensuing complex-dynamic structure formation (Section 2), that explains a remarkable similarity between the resulting brain and quantum structure behaviour, but the causally complete origin of dynamic complexity and the related complexity correspondence principle (see above) also shows that the microscopic, quantum world dynamics and brain function dynamics definitely belong to very *different* complexity levels (as opposed to numerous *directly quantum* brain models in the unitary theory [10-12,28-39]). The fact that the much higher level of brain complexity shows striking similarity to quantum system behaviour reflects the universal *holographic, or fractal*, property of the hierarchy of world complexity [15], where any well-defined system part tends to reproduce approximately the dynamical structure of the whole, but with proportionally *smaller* "resolution" (i.e. smaller number of features, or realisations, which just determines system complexity). Since the usual, dynamically single-valued theory and approach *cannot* see that dynamically multi-valued (probabilistic) fractal hierarchy of permanently changing system structure, it is obliged to evoke the single "acknowledged", but mysterious (unexplained) and *formally postulated* case of that kind of behaviour, i.e. that of a quantum system, in order to account for another, somewhat similar "miracle" of the unreduced dynamic complexity, that of intelligent and conscious brain operation (the same "quantum" mystification is used intensely by the same unitary science to account for various "miracles of life" and similar manifestations of genuine dynamic complexity in social life, see e.g. [39-42,45,50]). However, all the miracles of unreduced complexity, at any quantum and classical (including conscious) levels of world dynamics, *qualitatively different* among them, obtain their causally complete, i.e. totally realistic, consistent and intrinsically unified, explanation in terms of real, dynamically multivalued and fractal, interaction dynamics [15-24].

As we have seen before, eqs. (1), (2), (10), (38) are general enough to account for the above complicated electro-chemical combination of brain interactions (including interaction with the environment), and in particular, being expressed in terms of system element dynamics, they lead to the same, standard system of equations, eqs. (5) and (9). It would be convenient to consider that the separated degrees of freedom $\xi$ account for the more rigid, "chemical" degrees of freedom, including the initial system structure configuration (i.e. "mechanical"/spatial and related biochemical brain structure on a relevant scale), while one/several of the $Q \equiv \{q_i\}$ variables correspond to the "global" (inter-neuron) e/m patterns and other to the internal neuron excitations. The resulting state-function $\Psi(\xi,Q)$, eqs. (14)-(20), (26) (where the explicit time dependence is included in $Q$ variables), represents the entangled electro-chemical dynamical pattern of brain activity, accounting for all its functions. The most complete *general solution for the brain state-function* is provided by the universal, *causally probabilistic* and multi-level sum of eq. (26) over the emerging fractal hierarchy of system realisations, each of them obtained by the unreduced EP formalism, eqs. (11)-(17), (23)-(25) (together with respective values of dynamic realisation probabilities). As the analysis of the detailed realisation structure, eqs. (19)-(20), shows (see also [15-24]), the dynamically chaotic realisation change process at each level of dynamic fractality and within entire probabilistic fractal of brain activity pattern occurs inevitably in the form of the *generalised quantum beat* (essentially nonlinear, catastrophic self-oscillation), consisting of unceasing cycles of system *dynamic reduction (squeeze)* to the regular, localised realisation configuration it currently takes and the following opposite *dynamic extension* to a delocalised state of the *generalised wavefunction* (intermediate, or main, realisation), where the localised state (regular realisation) involves maximum *dynamic entanglement* of the interacting degrees of freedom (here the e/m and chemical constituents) and the delocalised state of wavefunction is obtained by the opposite *disentanglement* process, transiently "liberating" interaction components that perform the automatic dynamical "choice" of the next regular (localised) realisation.

If we take into account the *dynamically fractal (multi-level and hierarchically unified) structure* of the quantum beat pulsation and the *generalised, causally derived Born rule* for realisation probabilities, eq. (37), then we obtain a rather complete and unified picture of complex brain dynamics in the form of those unceasing, essentially nonlinear, global and fractally structured cycles of brain activity (as measured by e/m and chemical component density/flux). Due to its "omnipresent" and permanently changing structure at all scales, the generalised quantum beat solution explains the observed "binding", "awareness" aspects of intelligence and consciousness, while the fractally structured, detailed distribution of realisation probabilities on every scale according to the dynamic Born rule provides the causal, rigorously derived basis for the *meaningful brain operation* and the unreduced, "human" *sense* of the resulting information processing and *understanding*. In other words, the fractal system of centres of dynamic reduction within every global cycle of quantum beat pulsation is "automatically" (dynamically) concentrated around currently activated (functionally important) patterns of external (conscious and unconscious) "impressions", their processing, emerging "thoughts" and resulting "ideas". As those patterns change in accord with the "input data" or internal brain dynamics, the fractal structure of each quantum beat cycle automatically adjusts its probability (and thus density) distribution to system configuration, ensuring the *intelligent response* and conscious *understanding* (they are thus special, high-complexity cases of the universal *dynamic adaptability* of the unreduced complex dynamics, absent in any its unitary imitation [15-17]). In addition, the essentially nonlinear quantum beat of electro-chemical brain activity, as well as its internal fractal ramifications, gives rise to the *emerging internal time* (see Section 2.4 around eq. (30)), thus forming the physically real, universal basis for the necessary "sense of time" (internal clock) of an intelligent system that has nothing to do with the explicit time of eq. (38) originating from external (input) changes.

Although the global quantum beat pulsation (and their more localised manifestations) can be measured in the form of well-known oscillations of the brain e/m activity (see e.g. [1-9]), it is important to emphasize their essential and deep difference from any linear or even formally (but never *dynamically*) "nonlinear" oscillation models of unitary (dynamically single-valued and perturbative) theory. Indeed, the latter will not possess just those essential properties of truly autonomous emergence, flexible fractal "binding" of the entire brain activity and dynamic adaptability, which are especially important for understanding of consciousness (see also below). Another essential distinction from existing theories concerns the already mentioned generalised, "indirectly" quantum character of brain dynamics, which has only external, qualitative resemblance to the directly quantum dynamics at the lowest complexity levels and does *not* involve any microscopic quantum coherence on a nanometre scale and below (though the *real* similarity between these two well separated levels of dynamics has a rigorous complex-dynamic basis outlined above). Note also that high similarity between quantum (microscopic) and mental levels of complexity is due to the similar, predominantly "wave-like" character of the key entities at both levels (whereas this case is somewhat more different from "particle-like" level behaviour, such as that of "Newtonian" systems of permanently localised, rigid bodies). These results provide a *consistent* solution to persisting disputes around various "quantum brain" (and even quantum gravitation) hypotheses [10-12,28-40].

Recalling the universal Schrödinger formalism for the generalised wavefunction, eqs. (36), we find now that the wavefunction (intermediate realisation) of complex electro-chemical interaction dynamics in the brain, also designated as the *brainfunction*, $\Psi(\chi,t)$, satisfies the wave equation of the same kind, accompanied by the causally substantiated Born probability rule, eq. (37), that reflects (together with the causal quantization condition of eq. (35)) the unceasing dynamic collapses of the brainfunction to various regular (localised) brain realisations (constituting mental images, impressions, emotions, thoughts, ideas, etc.):

$$\mathcal{A}_0 \frac{\Delta \Psi}{\Delta t}\bigg|_{\chi=\text{const}} = \hat{H}\left(\chi, \frac{\Delta}{\Delta \chi}\bigg|_{t=\text{const}}, t\right) \Psi(\chi,t) , \qquad (39)$$

$$\alpha_r(t) = \alpha(\chi_r,t) = \alpha(\chi,t) = |\Psi(\chi,t)|^2 , \qquad (40)$$

where $\chi$ is the emerging regular realisation configuration, forming the new level of tangible space structure, or causally specified "mental space", made of neuron activities, thoughts and other patterns. The detailed structure of $\chi$ is obtained by dynamic entanglement of the interacting degrees of freedom $\xi, Q$ (essentially e/m and bio-chemical ones) according to the unreduced EP formalism, eqs. (11)-(15), (19), (20), (23)-(25). Similar to quantum-mechanical postulates (now causally explained themselves), the measured *dynamic probability*, $\alpha_r(t)$, of a brain activity pattern (*r*-th realisation) emergence is determined by the squared modulus of the brainfunction for that particular pattern, eq. (40), obeying the generalised, dynamically discrete Schrödinger equation, eq. (39). Note that similar to the mi-croscopic quantum mechanics, the Schrödinger equation for the brainfunction does not describe the quantum beat dynamics itself (i.e. system "quantum jumps" between regular realisations), but only the distribution of the probability amplitude (coinciding with the brainfunction density) for the emerging localised patterns (regular realisations): it is the result, rather than the origin or development, of the quantum beat process. Correspondingly, the time dependence in eqs. (39), (40) comes essentially from external interactions (within the Hamiltonian operator), rather than the emerging system time (hidden e.g. in the coefficient $\mathcal{A}_0$ in eq. (39)).

The Hamiltonian configuration expresses the pre-existing, "hardware" brain structure and can be approximated, in principle, by various model equations, unified e.g. within a series expansion of eq. (36b):

$$\mathcal{A}_0 \frac{\Delta \Psi}{\Delta t}\Big|_{\chi = \text{const}} = \sum_{n=0}^{\infty} h_n(\chi, t) \left(\frac{\Delta}{\Delta \chi}\Big|_{t = \text{const}}\right)^n \Psi(\chi, t) . \qquad (41)$$

The discrete form of differential operators in eqs. (39), (41) reflects the *dynamically discrete (or quantum)* character of unreduced interaction dynamics resulting from its *wholeness* [15,16] and appearing as visible discreteness of observed brain activity patterns. This kind of essentially nonlinear structure of unreduced brain dynamics, starting from the global quantum beat, may appear *externally* as a quasi-periodic pattern, but it is quite different from any unitary oscillation by its origin and internal dynamics. Nonetheless, at sufficiently fundamental levels of complexity or sometimes in the case of quasi-periodic behaviour (the limit of multivalued SOC, see Section 2) the discrete form of the dynamic equation for the brainfunction can be replaced by the usual, continuous version during its limited, "external" analysis.

There is, however, another important distinction of the universal Schrödinger formalism from any unitary model that can hardly be neglected, especially for the brain dynamics: the former implies, contrary to the latter, the unreduced, *dynamically multivalued*, and thus truly *chaotic, solution* (Section 2) that provides many essential, easily observable properties of the real brain operation (we discuss them below). This feature, as well as the entire complex-dynamic understanding and description of the brain dynamics, highlights the dynamically *emergent,* structure-forming, holistic character of any brain property thus derived, as opposed to various unitary imitations that *cannot* describe explicit structure emergence in principle and are forced therefore to *artificially insert* any its property with the help of a *postulated, mechanically fixed* structure or lower-level property. Just as the "miracles" of true intelligence and consciousness cannot be reduced "globally" to the postulated miracles of standard quantum mechanics (see above), their essential features cannot be consistently explained by various "local" models of neuron operation, such as the well-known "integrate-and-fire" model. Such models may only reflect particular details of individual neuron interaction acts, which can eventually constitute important features, but cannot directly account for the emerging result of *many* closely related individual interactions, permanently (and essentially) changing in time. Due to its inherent universality, the above brainfunction formalism and causal interpretation refer, in principle, to any level or scale of fractal brain dynamics, from the whole brain to any its level or activity pattern. In particular, the universal interaction complexity development (Section 2) will appear in the form of natural, generally irregular alternation of patterns of both limiting cases of complex dynamics, the more permanent (distinct) structures of multivalued SOC and irregularly changing (smeared) patterns of uniform chaos.

## 3.2. Emergent complex-dynamic intelligence and consciousness: Unified definition and properties

Having thus established the general dynamic content of neural networks with massively interacting components, we can now proceed with the dynamical meaning of the *emerging properties of intelligence and consciousness*. Already the obtained general picture of unreduced interaction development and complexity properties, applied now to the neuron interaction processes, show that intelligence and consciousness can only be understood as big and high enough *levels of unreduced dynamic complexity* (where the level of consciousness is generally higher than that of intelligence). In agreement with the general probabilistically fractal structure of complexity [15-17], complexity levels of neural network dynamics have hierarchical, fractal structure, where big enough "branches" (levels) describe qualitatively specific types of behaviour, separated by "steep" and big enough (but still physically continuous) complexity "jumps" from those of lower (and higher) levels of unreduced dynamic complexity. Since general (true) intelligence, including its unconscious, "animal" forms, is characterised by efficient control of a large enough environment, its (minimum) complexity level can be defined as that of the *complete environment complexity* (including the reverse influence upon it from intelligent species, etc.).[4] The necessary part of this condition follows from the complexity correspondence rule outlined above, while its sufficiency can be related to the "principle of parsimony" (Occam's razor), which can, however, be causally derived itself as another aspect of the same complexity correspondence principle. In other words, the dynamic complexity of intelligent behaviour can come exclusively from interaction of the intelligent system with its "generalised" environment and will therefore, in its *sufficient* version, only slightly (though definitely) exceed the total complexity of the latter. It is worthy of noting that contrary to lower-level dynamic complexity of non-intelligent systems (including living organisms) that can also quite "successfully" exist in the same environment, a truly intelligent system will *concentrate* within its *individual*, single "copy" the complete, *distributed* complexity of the dynamic environment.

In accord with our universal complexity definition (Section 2.4), this level of complexity, where the true intelligence begins, can be expressed quantitatively in terms of the number of permanently changing realisations of all interactions in the "generalised environment". However, it is the qualitatively big and high enough *level* of complexity that is much more important than the particular realisation number it contains (the latter can vary considerably during internal development of any given level of complexity), which explains why certain "minimum" *natural* intelligence can be defined rather well (although it inevitably has a "fractal", partially smeared structure), despite apparently large possible variations of various environment details. In this sense, if we define artificial (any) intelligence in a similar way with respect to any (artificial) environment complexity, it can certainly vary in a much larger range, including systems whose "perfect" intelligence in a particular, restricted environment will become totally useless ("nonintelligent") in another environment with higher complexity

---

[4] In fact, the highest complexity of any well-established (developed) environment is determined basically by its intelligent components (if any), which interferes self-consistently with intelligence definition as environment complexity and explains why the level of (minimum) intelligence depends relatively weakly on the details of nonintelligent environment dynamics.

(such situations can certainly happen occasionally also for natural intelligent systems).

Being a direct and "minimum sufficient" reflection of the environment complexity, the nonconscious intelligence is inevitably characterised by the globally chaotic kind of dynamics, as opposed to the limit of multivalued self-organisation (Section 2.3). Therefore such minimum, or animal, intelligence is qualitatively insufficient for appearance of the main properties of conscious behaviour.[5] The next higher level of brain dynamic complexity able to provide the minimum true consciousness is naturally obtained then in the form of simplest *permanently localised*, SOC type of structures, which can be realised as *elementary bound states* of nonconscious (but typically intelligent) brain patterns. At this point a general analogy with similar complexity development at its lowest, quantum levels can be useful. Dynamic consciousness emergence in the form of bound states is analogous to complex-dynamic emergence of the level of permanently localised, classical states from purely quantum, delocalised and chaotic behaviour at the lowest complexity sublevels [15,16,20-22]. If two elementary particles, such as proton and electron, form an elementary bound system, such as atom, then the probability of their simultaneous quantum jumps in *one* direction is low and quickly decreases with the number of jumps (in the same direction). This is because the quantum beat jumps of *each* of the bound particles are *chaotic* and *independent* from those of its partner. The bound quantum beat processes can therefore only perform their chaotic "dance" around each other, but cannot progress together to a big distance in one direction (in the absence of external force).

Now, the same mechanism of "generalised classicality" emergence in a bound system applies also to the emergence of localised, conscious states in the brain in the form of bound systems of various strongly chaotic, delocalised structures of unconscious levels of brain complexity. The first conscious level of brain activity results therefore from further (binding) interaction of unconscious activity products ("generalised impressions" from the environment) leading to formation of various *bound, permanently localised, or conscious, states* (their life time should be at least much greater than the period of internal quantum beat of each bound component). These simplest "elements of consciousness" start then interacting among them to form new localised (SOC) or globally chaotic states of higher sublevels, which constitute the developing structure of growing consciousness complexity. Such *additional* interaction with respect to unconscious intelligence needs a special "space" for its development and result accumulation, which explains the emergence and functional role of the *cerebral cortex* in the human brain as *inevitable* feature of conscious brain structure, where those bound, conscious states can form and further interact among them, giving rise to conscious "imagination" and similar specific features of *independent, internal* consciousness dynamics. Correspondingly, the unreduced complexity of conscious brain dynamics does not need to be limited any more to that of a particular environment and can grow to comprise and create ever new features of real or imaginary world. Note that similar to purely intrinsic, dynamic origin of classicality from quantum behaviour at the lowest complexity levels that needs no external, artificially imposed "decoherence" of the unitary theory, the complex-dynamic origin of consciousness results basically from internal brain interactions, using interactions with the environment only as a source of "input data" (fixed initially at the unconscious complexity levels). We see again that the analogy with quantum complexity levels provides a useful "holographic" reproduction of similar complexity development features, but does not imply the direct quantum (microscopic) origin of consciousness.

---

[5] This result is actually close to the conclusion that a natural environment in the whole cannot possess itself any kind of emergent, dynamic consciousness, irrespective of its detailed interpretation, while the same environment can, in principle, be characterised by a (nonconscious) intelligence determined by the highest complexity of intelligent species living in it (if any).

Consider in more detail the simplest case of a conscious structure emergence in a binding interaction of two nonlocal, globally chaotic unconscious structures. Each of them is represented by a *complex-dynamical quantum beat process* (essentially different from any regular structure or dynamics!) at the level of unconscious intelligence, characterised by unceasing change of $N_\Re \gg 1$ realisations taken in a dynamically random order (let $N_\Re$ be the same for both interaction participants, for simplicity of expressions only). The probability of a quantum jump of each of the interacting quantum beat processes towards any its particular, localised realisation is $\alpha \simeq 1/N_\Re$, in agreement with the general expression of eq. (17a). When the two interacting unconscious structures form a conscious, bound state, the probability of their correlated jump in one direction is $\alpha_{\text{corr}} = 1/N_\Re = \alpha \ll 1$ (whereas the probability of arbitrary jumps, or system existence as such, is evidently $\alpha_\text{arb} = 1$). Similarly, the probability of *n* consecutive jumps in one direction is $\alpha_n = (\alpha_\text{corr})^n = (1/N_\Re)^n = \alpha^n = \alpha^{\chi/\chi_0} = (N_\Re)^{-\chi/\chi_0} = \alpha(\chi)$, where $\chi = n\chi_0$ is the total distance of chaotic system wandering and $\chi_0$ is the length of elementary jump of each component, both expressed in terms of respective brain space coordinate $\chi$. We see that $\alpha_n = \alpha(\chi)$ decreases exponentially with $\chi$, so that the non-interacting bound system will remain localised within its size, of the order of $\chi_0$.

It is important, however, that the complex-dynamical "internal life" (*chaotic* realisation change) continues within such localised conscious state, ensuring its proper evolution in interaction with other, conscious and unconscious, brain states within the *unceasing* and *unifying* quantum beat dynamics. We deal here with an essential difference between the unreduced, dynamically multivalued self-organisation and its dynamically single-valued (unitary) models in usual theory. It explains, in particular, why the dynamics of consciousness is characterised by much slower processes than unconscious reactions: according to the universal criterion of absence of global chaos, eq. (28b), the system should be far from its main resonances in order to preserve a distinct enough, e.g. localised, configuration and changes, and therefore, at its slow component rate. The role of chaoticity/complexity is also reflected in the above expression for $\alpha(\chi)$ showing that localisation grows with $N_\Re$ and disappears at $N_\Re = 1$, i.e. in the (unrealistic) case of single-valued, regular (or "averaged") dynamics with zero complexity.

It is not difficult to outline further brain complexity development within its conscious activity. It is important that each qualitatively new level of complex brain dynamics as if *starts from the beginning* in the image of the environment complexity it provides. Thus, conscious world reflection in terms of permanently localised elementary structures starts representing the same outside world dynamics that has already been properly reflected by unconscious levels of brain dynamics, but now acquires a "new life" in the form of permanent and subjectively "controlled" images of real entities, which become relatively independent of their real prototypes (especially for higher levels of consciousness). When this new, conscious representation of reality approaches a correct enough image of the external dynamic complexity, it naturally tends to produce a *general image of itself*, appearing as a state of *awareness* and giving rise to possible *next level of consciousness*. This superior level of consciousness operates now with indirect images of world complexity from the first level of consciousness, closely entangled among them in a system of holistic "associations". This superior consciousness "looks" upon its own complex-dynamic (and generally localised) images of external dynamical patterns, at least as much as at those patterns directly. We obtain thus the detailed complex-dynamic interpretation of the property of *reflection* of conscious brain activity. Since the genuine "technical" capacity of a large neural network is fantastically high [16] (see also below), far beyond usual unitary estimates, the hierarchy of complex-dynamic reflection levels can grow considerably to ever superior levels of consciousness, where already the lowest level provides the necessary minimum for conscious understanding of the environment.

The emergent, complex-dynamic consciousness is not only explicitly obtained as a result of unreduced interaction processes in the brain, but possesses a hierarchic, multi-level structure, where each next level provides a qualitatively new, "superior" image (and extension) of reality, including complex-dynamic images of all lower levels of consciousness. Practical emergence of a new complexity level needs the suitable stock of latent interaction complexity, or dynamic information (Section 2.4), and is accompanied by a dynamic resistance (generalised *inertia*) of the already existing structures, so that the appearance of a new, big enough level of consciousness has the properties of a revolutionary change, or "generalised phase transition" [15]. The persisting qualitative difference between unitary and unreduced (complex-dynamic) reality images in (conscious) knowledge provides a relevant example of different levels of consciousness.

With this general dynamic picture of intelligence and consciousness and their internal development, let us verify now how exactly can it reproduce the known *properties of intelligent and conscious behaviour* (e.g. [1-9]), including those that can be postulated as necessary, empirically based demands for *artificial consciousness* systems [13,14].

Note, first of all, that our complex-dynamic interpretation of intelligence and consciousness provides their *well-specified* origins and definitions, including a clear-cut *distinction* between these two "close" levels of higher brain activity, remaining rather ambiguous within unitary approaches to both their natural and artificial versions.

We can proceed with the property of *autonomous dynamic adaptability (I)*, being common for intelligent and conscious reflections of reality. As we have seen above, this feature emerges as a universal property of any unreduced, complex interaction dynamics (absent in its unitary imitation), while its necessary magnitude for the efficient intelligence and consciousness is determined by the complexity correspondence principle relating the degree of adaptability with the sufficient dynamic complexity of the brain that should exceed that of the controlled environment (we provide quantitative estimates below).

The *"logical", "binding", and "supervising" features (II)* of a *conscious* system are obtained within the key interpretation of conscious states as physically *bound* states of chaotic quantum beat processes of electrochemical interactions in the brain neuron system, emerging as *localised* realisations of the *whole* system of brain interactions at a special complexity level, exceeding and therefore *including* all realisations from lower, unconscious reflection of the environment.

This superior structure of the level of bound conscious states underlies also all versions of clearly recognised *separation between the "self"*, represented by those *dynamically unified* bound states in the cortex, *and the "rest" (environment)*, the latter being reflected already at the lower level of unconscious intelligence *(III)*.

The superior, ultimately emerging form of this property is provided by the complex-dynamic *awareness (IV)* described above, where the bound conscious images of reality include that of oneself, i.e. the cumulative image of the conscious representation of the environment, actually forming the *next higher sublevel of complexity*. In terms of human species evolution (and for illustrative purposes only), property (III) could be figuratively designated as *Homo habilis*, while its version (IV) would correspond to the true (and still apparently uncertain) *Homo sapiens*.

*Practical abilities* of a conscious brain (also present, in a reduced form, at the level of unconscious intelligence), such as *reality control and self-control, imagination and planning-anticipation (V)*, follow from the emergent, interaction-driven origin of the corresponding brain structures, where higher-level conscious, bound structures acquire their own dynamics (in principle of ever growing complexity), showing only general, weak dependence on the environment.

The properties of intelligent and conscious systems summarised as *emotions, desires and motivations (VI)* are manifestations of universal *creativity* of complex dynamics expressed by the universal symmetry (including transformation) of complexity (Section 2.4) and appearing also as "élan vital" in the development of any living system: it is a result of interaction potentialities expressed by the dynamic information and forced, by the *unreduced* interaction itself, to develop into the fully unfolded system structure, or dynamic entropy.

Finally, the *sustainable, autonomous growth of intelligence and consciousness* underlying also the property of *education/learning (VII)* results from the same complexity development of the unreduced interaction process, constituting thus the basis for unlimited (in principle) growth of consciousness, as explained above.

In accord with the complexity correspondence principle, any of the above properties (I)-(VII) of the dynamically multivalued, essentially nonlinear and intrinsically creative interaction processes in the brain neuron system *cannot* be properly reproduced by conventional, unitary theory, just because of its dynamic single-valuedness and strictly *zero* value of unreduced dynamic complexity, which is the unified, genuine origin of all difficulties and ambiguities in the existing understanding of consciousness, irrespective of details [1-11]. Indeed, the unitary reduction of real interaction in the canonical theory cannot explain even the *simplest*, quantum system behaviour at the *lowest* complexity levels and is forced to postulate the "impossible" and "inexplicable" properties of those *real* systems in the form of "quantum mysteries" and "paradoxes" (see [11,15,16,20-24] for more details). This intrinsic deficiency of unitary theory is inherited by its complexity imitation at higher levels of world dynamics. Therefore the existing "general" applications of those effectively zero-dimensional imitations from the unitary "science of complexity", often in a characteristic post-modern "hermeneutics" style, can create essential confusion in the already quite obscure field of knowledge. Speculative description of consciousness in terms of "attractors" and other abstract "models" of unitary theory (see e.g. [51,52]) operates, in fact, with zero-complexity entities and is unable to explain even much simpler structures than those of conscious brain dynamics. An "attractor" is produced by a *continuous trajectory* of a system with *fixed, postulated configuration* in an *abstract, artificial*

"space" and therefore has nothing to do with the real system dynamics based on the permanent and qualitative change of its configuration, obtained as inevitable, generic consequence of the unreduced dynamic equation solution (reduced to a trivial change of notations in the conventional, effectively zero-dimensional, perturbative "approximation"). Replacement of *dynamic, interaction-driven* multivaluedness of *incompatible* system realisations (its *different configurations*) and *probabilistic fractality* by "multiple attractor basins" produced by a *postulated* system configuration and *coexisting* in an abstract space is a very rough verbal trick of the unitary imitation of complexity, which *cannot* explain *any* property of the unreduced system dynamics, but persists nevertheless in many "serious" sources on interpretation of its highest-level property, consciousness.

The huge contrast between the unreduced, multivalued dynamics of a multi-component interaction system and its unitary projection appears in a yet more transparent form within a quantitative estimate of the total brain power [16]. The unreduced power of a complex-dynamic process, $P$, i.e. the maximum number of operations it can perform per time unit or the number of units of information it can store, is proportional to its dynamic complexity $C$ as given by the full number of regular realisations $N_\Re$: $P = P_0 C(N_\Re) = P_0 N_\Re$, where the coefficient of proportionality $P_0$ is of the order of the unitary, sequential operation power, so that the relative power of unreduced, complex-dynamic process is given by its realisation number, $\delta P = P/P_0 = N_\Re$. If our natural or artificial brain consists of $N_{\text{cell}}$ "generalised neurons", each of them connected in average to $n_{\text{link}}$ other cells, then the total number $N$ of system links is $N = N_{\text{cell}} n_{\text{link}}$. The distinctive property of the unreduced, multivalued system dynamics is that the total realisation number is given by *all possible combinations of links*, i.e. $N_\Re \simeq N!$, whence

$$\delta P = N_\Re \simeq N! \simeq \sqrt{2\pi N}\left(\frac{N}{e}\right)^N \sim N^N, \qquad (42)$$

where we have used the Stirling formula valid for large $N$. Since for the human brain we have $N_{\text{cell}} \sim 10^{10}$ and $n_{\text{link}} \sim 10^4$, the estimate $N = 10^{12}$ for the number of conscious brain links should not be exaggerated. The expression of eq. (42) gives for $N = 10^{12}$ the following estimate for the relative power of complex-dynamic brain operation: $\delta P \gg 10^{10^{13}} \gg 10^{10^{12}} \sim 10^N$, which is a *practical infinity*, meaning that the real, dynamically multi-valued brain power is "infinitely" greater than that of its unitary, mechanistic models.

This "astonishing" result is certainly due to the *complex-dynamic parallelism* of the unreduced interaction dynamics, where the system itself creates, in a real-time mode, the necessary dynamic structures and ways of search for a solution. The mechanistic "parallel information processing" does not have this property and represents only additive reconfiguration of the same sequential dynamics that cannot provide a real gain in power (with the same "hardware" capacities). Indeed, assuming that the average frequency of brain realisation change is not less than 1 Hz (which is a very moderate estimate), one can compare the above estimates of complex-dynamic brain power with the unitary estimate of the "ultimate" computation power for the whole (known) universe [53] to see that the former remains "infinitely" greater than the latter [16] (although curiously this unitary estimate of the power of a very special, "quantum" computation process relies on a strong emphasis of "advanced", "magic" parallelism and "complexity" [54], demonstrating once more the absence of any power in unitary imitations of complexity). The inevitable payment for such tremendous superiority of the unreduced complex-dynamic power takes the form of irreducible dynamic randomness, just underlying the above huge efficiency. However, the related uncertainty of result is not really a problem, since it can be reduced to a necessary minimum in the multivalued SOC regime, without any essential loss of the total operation power. It is easy to see that the huge values of $\delta P$ provide a quantitative expression of the "magic" qualitative properties of complex brain operation, such as those of intelligence and consciousness [16]. This conclusion will remain valid for much smaller values of $N$ that can be expected for artificial neural networks, thus underlying the corresponding "magic" power also for artificial intelligence and consciousness, produced by their *unreduced*, complex (multivalued) dynamics.

# 4. Complex-dynamic machine consciousness and its social implications

Since the above causal understanding of consciousness is based on the unreduced analysis of a real, full-scale system of interacting elements explicitly producing the full version of this property, it can form the *truly scientific*, fundamental and rigorous, basis for the machine consciousness concept replacing today the previous paradigm of artificial intelligence (Section 1) [13,14]. Without rejecting possible more limited, dynamically regular versions of machine consciousness, we shall consider now specific features of its unreduced, complex-dynamic (multivalued and chaotic) realisation in the artificial system of connected elements (neural network), as well as various technological, social and mental implications.

Note, first of all, that the proposed unreduced, complex-dynamic version of conscious control and information systems should be considered as the *inevitable*, qualitatively new and *already urgently needed* stage of development of modern technology. Indeed, if we apply our *universal* description of various dynamic regimes of arbitrary systems with interacting elements (Section 2) to modern technological systems, we conclude immediately that their operation refers basically to the limiting, almost regular regime of multivalued self-organisation. However, as the compositional (configurational) sophistication of technological (and related social) systems inevitably grows, it finally and *inevitably* attains a level, where the unreduced interaction complexity and related true chaoticity will appear explicitly, with a high enough magnitude. This conclusion follows from the rigorous *criterion of chaos* of the universal science of complexity, eq. (28a), showing that one can avoid explicit, big chaoticity only by maintaining the system far away from all its essential resonances. But as the technical (and socio-economic) intricacy of the system grows, its resonance conditions inevitably get closer, so that one *cannot* avoid their overlap and thus essential chaoticity above certain critical intricacy of system composition. Needless to say, this critical level of global technical and social complexity is being exceeded by modern globalised technology in a growing number of cases. Since the unitary, regular technology and society paradigm practically rejects those chaotic elements, they inevitably appear in the form of undesirable, more or less catastrophic (propagating) system *failure*, which should be compensated by more and more frequent and inefficient direct, "extra-ordinary" human interventions in otherwise automatic processes. Our analysis shows that there is no other issue from this growing "crisis of complexity", than explicit "acknowledgement" of the unreduced, *really existing* dynamic complexity (multivaluedness) of technological interaction processes, followed by transformation of its "destructive" influence on the *unitary* control scheme into huge advantages of its constructive, unlimited realisation outlined above. This result and approach applies to *any* level of technology, economy and social life [15,55-57], but we shall concentrate here on the highest levels related to conscious control systems.

Such truly conscious technical systems of control and communication can possess the *genuine*, complex-dynamic consciousness property (Section 3.2), but which at the same time *differs* essentially from the natural, human version of consciousness by the characteristic *shape* of the conscious operation complexity. As we have seen above, the true consciousness emerges as certain, high enough level of dynamic complexity, which exceeds considerably those of arbitrary living and intelligent systems, having very high positions themselves in the hierarchy of world dynamic complexity [15-17]. In the case of natural consciousness, i.e. the one obtained within a "natural" (biological) evolution, this means that carriers of consciousness should "first" be alive and then intelligent in order to have a (generally rare) chance to develop at least a minimum level of conscious intelligence. Since the lowest level of "intelligent" complexity is determined by the maximum environment complexity (in a reduced formulation, the same will be true for any living system complexity), it follows that the natural consciousness structure, at its initial, lowest levels, is inevitably characterised by a specific, relatively *flat* shape of its internal hierarchy of complexity (dynamical fractal) resembling a "pancake". Whereas the "consciousness pancake" should have a minimum thickness corresponding to the lowest level of consciousness complexity, its relatively large width is inherited from the shape of unconscious intelligence and comprises a high diversity of the controlled environment complexity (though represented by properly localised, conscious states of brain activity, see Section 3.2).

By contrast, *artificial*, man-made systems of machine consciousness need not and actually should not incorporate the entire "horizontal" diversity of a living environment complexity, but do need to have a minimum

"vertical" dimension of the localised (SOC) reflection of their *limited* environment. Therefore the systems of artificial consciousness emerge in the shape of relatively narrow vertical "rods" (or other "pyramidal" structures) in the "space" of universal hierarchy of complexity. They will have the "air" (characteristic behaviour) of very narrow, but highly qualified, conscious "specialists" in their particular environment, knowing very much about it, but very little beyond it, and therefore suddenly becoming very "stupid" just outside of their "professional interests" (this phenomenon is known, in its milder form, also for human consciousness realisations). Let us emphasize once more that such limiting, "vertical" shape of a carrier of consciousness complexity would be *impossible* for any truly natural, living system, but can and should be realised for (truly) conscious machines, providing a *fundamental*, rigorously substantiated basis for their creation and making the latter much more *realistic* (as opposed to an ill-defined imitation of the full human consciousness). Therefore we can propose this conclusion as a *well-specified scientific basis* for the *concept and paradigm of artificial, but genuine consciousness* (and thus also *intelligence*), including its *rigorously derived definition* in terms of the above *level and shape of unreduced dynamic complexity*.

Taking into account the complex-dynamic machine consciousness concept thus specified, we can further advance towards *scientifically rigorous* understanding of *social* and *mental* implications of artificial consciousness by considering the next interaction level between conscious machines and natural consciousness carriers. It is easy to see, for example, that practical, professionally intense interaction between "complexity rods" of conscious machines and "pancakes" of minimum levels of natural consciousness provides an efficient way of otherwise difficult development of natural consciousness towards its higher-level, less flat shapes, where multiple "rods" of artificial consciousness would "impose", at least partially, their "vertical" dimensions to a naturally diverse, but vertically limited consciousness of living beings. In other words, interaction with (truly) conscious machines can become a very efficient, and quite possible the only real, way of *massive* natural consciousness development (otherwise stagnating or even turning into degradation).

A complementary conclusion, following from the complexity correspondence principle (or the underlying symmetry of complexity), states that systems of artificial consciousness cannot exceed the level of consciousness (complexity) of their creators, which in our case are assumed to be carriers of natural consciousness. At this point we switch from mental to social aspects because it follows that the *only* consistent dynamics of progressive (complexity-increasing) society development can result from (massive) interaction of its lower-consciousness members with conscious artefacts produced by efforts of members with (essentially) higher consciousness level (if any). One consequence is that society, which does not contain members with big enough difference of their consciousness levels or cannot profit from it by efficient interaction, is unable of (internal) complexity development and therefore condemned to disappearance: any other interactions (e.g. using only zero-complexity machines of unitary technology) *cannot* provide complexity growth to higher levels of consciousness. It is impossible not to note that this rigorously derived conclusion directly contradicts the currently dominating egalitarian social doctrine of a "democratic" flavour (often trickily exploited). On the other hand, one may argue that interaction between higher- and lower-consciousness society members can proceed by their direct, "natural" communication, including science, education, etc. However, real-life experience clearly demonstrates too low, "subcritical" efficiency of such "natural" interaction, even in the best cases, which is additionally hampered by *inevitable* intervention of machine-intermediated interaction within a technically developed civilisation, where the unitary, zero-complexity machines *impose* their *ultimately low complexity* to the entire system of strong mental and social interactions and its results. In this situation the qualitatively new, complex-dynamic, intelligent and conscious machinery can be the *only realistic*, and actually very strong, *catalyst* of natural consciousness development within a machine-based civilisation, underlying its development in the whole [15].

Inspired by this great purpose of the genuine machine consciousness paradigm, we can turn now to discussion of practical details of its realisation, following from the above description (Sections 2 and 3). Since in principle there is no problem today with fabrication of elaborated enough networks of connected elements ("neural networks"), the specific features of conscious networks involve their detailed structure and imposed operation modes. The general conclusion of our analysis implies that the true, complex-dynamic intelligence and consciousness can appear only in a sys-tem with high enough *freedom* of interaction between elements that *cannot* be based on pre-programmed, regular interaction rules and detailed results as it occurs for all unitary machines. Any detailed programming of regular interaction details should be abandoned in the case of complex-dynamical devices in favour of their natural, dynamic complexity development (including their interaction link modification), though occurring in a general direction determined by the *universal symmetry of complexity* and the ensuing particular laws (Section 2.4).

A more specific result of the above consciousness analysis (Section 3.2) implies that intelligent/conscious system interactions cannot be reduced only to local "rapid" (electric) connections between individual elements, but should also include a complementary distributed, dissipative, "slow" component necessary for efficient dynamic unification and stability of artificial brain dynamics in the form of generalised quantum beat. In the natural brain such component is provided most probably by chemical neuron structure and interactions, but such "bio-inspired" construction of artificial conscious systems may be not the easiest one. Another candidate for that "slow" interaction component is provided by properly configured magnetic materials and interactions, "repeating" generally (but not exactly) the electric connection interface and interacting with it "almost everywhere". It is not difficult to see that the detailed realisation and principles of construction of such explicitly complex-dynamic networks will be very different from the now realised unitary approach and technology, but as we have shown above, this way of development is objectively inevitable and unique at its next stage starting already today (see also ref. [16] for similar results for the unreduced nanotechnology concept).

In conclusion, we would like to emphasize once more the far-going mental and social implications of the genuine artificial consciousness paradigm, which have been briefly outlined above and would certainly need further development using this intrinsically interdisciplinary approach and our universal dynamic complexity concept and formalism. The general motivation for these studies is as big as civilisation development in the whole, since the above rigorous analysis shows the indispensable, unique role of complex-dynamic (multivalued) interaction processes and technology for progressive civilisation development today (see also [15,16,55-57] for the universal concept of development). Since artificially produced, technical structures play a major and ever growing role at any scale of world development that *cannot* be abandoned or turned back, increasing replacement of their currently dominating, complexity-suppressive design and operation mode by the unreduced, *explicitly complex-dynamic technology*, inevitably comprising *key elements of genuine machine consciousness*, emerges as the objectively substantiated, *uniquely* progressive way of development, including creative progress of *individual* natural consciousness as its *inherent component*. Note finally that the use of much more restricted, unitary versions of machine consciousness, which can only imitate, but not reproduce the unreduced consciousness features (see their list in items (I)-(VII) in Section 3.2), can be considered as a first-step motion in the same direction of growing complexity, which should not replace, however, the search for and practical realisation of explicitly complex-dynamic, truly intelligent and conscious machinery.

# 5. Crisis in science, complexity revolution and the transition to intrinsically sustainable civilisation

The discussion of the complex-dynamic intelligence and consciousness concept in previous sections revealed deep human and social implications of the consciousness problem, far beyond the issues of purely scientific understanding or concrete applications. Moreover, the rigorous analysis of the modern world condition within the same, universally valid concept of dynamic complexity shows that the entire system of human civilisation has just entered, in these recent years, into a *global critical state*, after which it can either continue to a quick and fatal decline, within the current tendency, or perform the crucial transition to higher-complexity and higher-consciousness level, where it can start the new, now fundamentally unlimited, or truly sustainable, progress [15,16,55-57]. It is important to recognise the intrinsic, well-specified relation and qualitatively new, nontrivial content of those key issues of the unreduced dynamic complexity and its *necessary revolution* in *both* scientific understanding of reality (dynamic multivaluedness paradigm) *and* the effective general level of social and individual consciousness.

In following sections we specify the details of interrelated major aspects of this necessary complexity transition, including "the last scientific revolution" towards the unreduced, universal dynamic complexity paradigm (Section 5.1), the complexity revolution and transition to a superior level of consciousness on a broader scale of entire society and global civilisation (Section 5.2) and the targeted resulting condition of truly sustainable progress, with its concrete purposes, features and criteria (Section 5.3).

## 5.1. The end of unitary science and the beginning of causally complete knowledge

Modern deep crisis in fundamental science becomes ever stronger and more evident on a growing scale of fields and aspects (see e.g. [15,16,57-63] and further references therein). It is the extreme, limiting parts of the latter that reveal especially striking and quickly accumulating signs of a critical state, including the persisting old and growing new "mysteries" in the microworld of particles and fields, the glaring contradictions in cosmology on various astronomical scales (up to the entire universe), and the ultimate complexity of living matter, conscious brain and developed society dynamics. It is no coincidence that this critical state of fundamental knowledge reproduces the critical bifurcation of the embracing civilisation development, mentioned above, with its ultimately painful choice between the default fatal degradation and unlimited new progress at the superior level of conscious understanding.

Note the key difference of such kind of real crisis in modern science from its much more comfortable description within the well-known concept of the end of science [59] as being due to a "practically perfect" (or at least "fundamentally saturated") state of knowledge that does not reasonably imply any essential further progress. This latter vision shows a strange correlation with the equally deficient (and much more widely accepted) attitude towards the state of the encompassing social system representing its current (unitary) "liberal democracy" realisation as close to a practically possible (positive) maximum of efficient social order. In reality, both modern official systems of knowledge and social organisation show visible "saturation" signs only within certain, actually very rough and artificially imposed simplification of complex reality, revealing huge and quickly growing contradictions, especially just when they have entered now into a critically unstable state of global bifurcation (with the ensuing inevitable transition to a qualitatively different state).

As *rigorously* shown in this and other related papers [15-24,55-57], this truly critical state of modern science and the related contrast choices for further development of entire civilisation have in reality a *well-specified, scientifically nontrivial origin* and *resolution* in terms of the *necessary transition* from the artificially limited, dynamically single-valued, or unitary, science paradigm, method and vision to the dynamically multivalued, causally complete content of the universal science of complexity. As a matter of fact, the absolutely dominating, dynamically single-valued "model" of the unitary science and world vision represents the *maximum possible limitation* of a great number of system (any real interaction process) realisations to just one, arbitrarily fixed and "postulated" realisation, corresponding to strictly zero value of the unreduced dynamic complexity (including all the cases of standard unitary "science of complexity") [15-24,55-57]. This unitary, effectively zero-dimensional (point-like) projection of reality could give an impression of correct description only for actually quite exceptional, "almost regular" cases of the limiting SOC regime (Section 2.3) exclusively considered in the "Newtonian" science paradigm, but even in such cases there are the well-known "unsolvable" (or badly solved) problems of genuine chaoticity, irreversibility and time origin.

This drastic simplification of reality corresponds also to the so-called "positivistic" ideology of unitary science (in particular, taking the form of "mathematical physics" in its most "exact" branches), which explicitly rejects the necessity of genuine, "truly complete" understanding of the observed reality (defended especially by René Descartes, the famous precursor of Newtonian positivism) in favour of its "model", simulating description, where inevitable glaring contradictions and ruptures (nonuniversality) are "compensated" by simplicity of an "exact" (or perturbative) model, "sufficiently correctly" imitating a particular kind of observable behaviour (but usually not other ones described by separate "models") and not resulting from the complete (interaction) problem solution. It becomes clear now that such only slightly decorated empiricism of the dominating positivistic ideology of the unitary science is but a general expression of the underlying dynamic single-valuedness, whereas the dynamic multivaluedness of the truly complete (and thus truly exact!) problem solution (Section 2) leaves no place for any "mysteries", ruptures and other contradictions of the postulated unitary "models". We can say therefore that the current "end of science" represents indeed the definite end of just that, actually very special kind of positivistic, or unitary, science, which should now be extended, with the help of the causally complete, universally applicable problem solution, to the intrinsically non-contradictory, at least locally complete knowledge of the universal science of complexity (it is limited only by inaccessible observation possibilities, which does not really change the practical completeness of the obtained picture of reality).

Whereas the rigorous basis of the universal science of complexity is properly exposed in Section 2 and its advantages in application to the problem of consciousness are revealed in Sections 3 and 4, we can emphasize here the general, "ideological", fundamental distinctions of this new, extended science framework with respect to usual, unitary science postulates and results [57].

(1) The entire framework and content of usual, dynamically single-valued, positivistic science is deeply based on the tacit *self-identity postulate* and related "rigorous" statements of various *theorems of uniqueness of problem solution*. The self-identity postulate implies the apparently "evident" identity of any mathematical (and respective real) structure $\mathfrak{A}$ to itself, $\mathfrak{A} = \mathfrak{A}$, including its formal time dependence, if any, $\mathfrak{A}(t) = \mathfrak{A}(t)$. However, as we have seen above (Section 2.1), the unreduced interaction process within any realistic structure has the dynamically multivalued result, in the form of multiple system realisations, *permanently replacing each other* in a causally (and truly) random order. Therefore any realistic structure, described in the causally complete framework of the universal science of complexity by the unreduced, dynamically multivalued problem solution, is *not* (exactly) self-identical, $\mathfrak{A} \neq \mathfrak{A}$, which also gives rise to the real, naturally unstoppable and irreversible *time flow*, $\mathfrak{A} \neq \mathfrak{A} \Rightarrow \mathfrak{A}(t)$ (Section 2.4). We deal here with the deeply diverging character of reality vision and presentation in unitary science and universal complexity science: the static, fixed structures of the former (including formal, postulated time dependence) are opposed to permanently, intrinsically and chaotically changing structure-processes of the latter (including *externally* regular or static structures and processes). As to the notorious "uniqueness theorems" of unitary science, they just realise a standard logical trap, where the tacitly presumed single-valuedness (e.g. of interaction potential) is then restated as the "rigorously derived" theorem conclusion [57]. As a matter of fact, no real, truly rigorous and complete problem solution can possess the uniqueness property, already because one cannot stop the real, entropy-increasing (and thus structure changing) time flow. Needless to say, already the most obvious empirical properties of intelligence and consciousness provide a clear demonstration of the absence of self-identity and uniqueness, in correlation with the persisting difficulties of genuine understanding of intelligence and consciousness within the unitary science framework [1-12].

(2) Another characteristic, inevitable property of the unitary projection of reality is its *absence of well-specified, tangible material quality* of purely abstract structures used (forming the very basis of the major "mathematical physics" approach), which is directly related to its artificially limited, dynamically single-valued scheme, or "model", of reality. As shown in our unreduced interaction analysis (Section 2.2), the missing material quality is rigorously defined in the universal science of complexity by the *dynamically multivalued entanglement* of interacting system/process components, organised into the hierarchy of *dynamically probabilistic fractal*, which is quite different from usual, unitary fractality and provides the truly exact (causally complete) description of real structures. It underlies also such universal properties, especially important for intelligent and conscious behaviour, as autonomous *dynamic adaptability* and *teleology* (creation of and evolution towards local and global purposes).

(3) The *absence of the origin of genuine randomness* in the dynamically single-valued approach of conventional science is closely related to its self-identity property and unique-solution projection (item 1). Whereas the intrinsic and omnipresent dynamic randomness and chaoticity is immediately implied by the unreduced, dynamically multivalued problem solution (Section 2.1), the unitary theory is forced to introduce its "widely accepted" notion of chaos and randomness in an artificial and contradictory way, in the form of "exponentially diverging" (but in principle regular) trajectories, incorrectly extending the perturbation theory result and supposed to "amplify" small, but already present, externally inserted "random deviations" in initial conditions. The same kind of incorrect trickery is used in other ideas of chaoticity in the unitary "complexity science", such as "strange attractors" or "routes to chaos". Note that only genuine, intrinsic and interaction-driven randomness is necessary for the observed efficiency of major searching, estimating and planning activities of intelligent and conscious systems.

(4) The *absence of genuine, dynamic discreteness* in the dynamically single-valued analysis is actually related to the problem of structure formation, or emergence, as such, fundamentally missing in usual theory and only artificially inserted into its models from the beginning (including the formally imposed space and time concepts and variables). While the discreteness of emerging real structures is provided, in the unreduced interaction analysis, by the dynamic discreteness of changing realisations and their complexity levels (Sections 2.1 and 2.4), it is replaced either by smooth unitary continuity or by false, mechanistically imposed discreteness in unitary science. Qualitative limitations and deep contradictions of these imitations are characteristic also of other interconnected mathematical constructions of standard theory, including the unitarity itself, calculus, evolution operators, symmetry operators, any unitary operators, Lyapunov exponents, path integrals, and statistical theories. Therefore all the related concepts of emergence, creativity and qualitative transitions omnipresent in intelligence and consciousness dynamics ("understanding", "ideas", etc.) can be consistently formulated only within the unreduced, dynamically multivalued description of the universal science of complexity.

It is important to emphasize finally the unique, *unifying global features* of the unreduced complexity description appearing as unified manifestations of the *single law*, the absolutely *universal symmetry of complexity*, which expresses the behaviour of the equally *unified world structure* in the form of *dynamically probabilistic fractal* (Sections 2.2 and 2.4). One can compare these rigorously derived and variously confirmed results [15-23,55-57] with ever increasing doubts about the very possibility of such kind of unification (or even about the existence of objective and truly fundamental scientific laws in general) [59,64-67], even within the limited scope of fundamental physics (after so many failed attempts to obtain it in various schemes of unitary "mathematical physics"). The comparison reveals also the astonishing blindness and deafness of unitary science practice, which is closely related to its strongly limited content [16,57].

## 5.2. Complexity revolution as the necessary transition to superior level of consciousness

As rigorously demonstrated in previous sections (see especially Sections 4 and 5.1), both techno-social demands of further civilisation progress and modern deep problems of scientific knowledge development necessitate the transition to a superior level of reality understanding equivalent to a higher level of (general) consciousness. According to the complexity correspondence principle (Section 3.1), this is indispensable already for the genuine, causally complete and practically efficient understanding of the phenomenon of consciousness itself, but actually appears to be necessary for further progress in all fields of fundamental science (and thus eventually technology and everything else), starting already from the lowest complexity levels (elementary particles and fields) [15-24,55-57].

One should speak therefore about a much deeper and wider transition in the entire civilisation development, implied by the universal science of complexity and the observed development tendencies, than the corresponding professional knowledge progress itself (which has, of course, its own importance and plays the key role in this transition). We deal here, in fact, with the ensuing "new way of thinking" and related new, complex-dynamic approaches in all spheres of human activity, as opposed to the still dominating tendency of simplification, even despite the evident advent of superior complexity in practical life (due to the previous huge, but mainly empirically obtained progress). It becomes obvious that the necessary transition from the traditional unitary tendency of maximum possible simplification (where informally "the simpler the better") to the opposite, exclusively progress-bringing tendency of growing complexity (where greater complexity is generally better) can occur only in the form of a highly nonlinear, rapid and collective enough (though fundamentally individual) "revelation" called here the *revolution of complexity* (or *complexity transition*) [15,16,55-57]. In particular, it is the *provably single possible way* to realise the notorious purpose of *sustainable development* (see also the next section), which remains but a strongly popularised case of wishful thinking (or even a dangerous illusion) within the traditional, simplification-based unitarity and its mechanistic approach, irrespective of the quantities of technical efforts applied (including the pseudo-"green" technologies, only imitating ecological advantages).

This latter case of sustainable development demonstrates an important general constituent and result of the imminent complexity revolution, which implies not only new ways, means and instruments, but also and especially the *new general purpose* of development and human life on every scale. It is evident that today such suitable, universal and practically sustainable purpose of civilisation development (in both its social and individual aspects) is persistently absent, probably for the first time in its "modern" history ($\sim$ AD). Former versions of the general purpose (under various religious, ideological and techno-scientific guises) have disappeared starting approximately from the beginning of the twentieth century (the famous religious-philosophical "death of God"), while no new ones, of a suitably high level, have appeared until now. The reason for this persisting (fundamentally) *purposeless* existence is precisely the necessary qualitatively big progress of consciousness in the direction of explicit complexity growth and efficient monitoring.

Now that the necessity of complexity transition becomes evident, together with the key implication of qualitatively growing, essentially complex-dynamic consciousness, we can formulate the expected new general, ultimate purpose at the forthcoming superior level of complexity-driven progress as the permanent, maybe uneven but unstoppable *progress of that unreduced, now practically unlimited human (including artificial) consciousness* (in all its now unified aspects of mind, spirit, emotion, etc.). It naturally starts with the key, step-like turn of complexity transition (from the still dominating unitary thinking) and then continues as a more gradual and stable, never-ending growth of consciousness dominating in all spheres of human activity. The huge, unlimited scale of this new general purpose, far beyond any traditional limitations and separations of usual "science", "spirituality" and "practical life", demonstrates the true extension of consequences of the above dynamic multivaluedness and related universal complexity concept (Section 2), without any decrease of its intrinsic scientific rigour (taking into account, in particular, its causally complete concept of consciousness, Sections 3 and 4).

## 5.3. Genuine sustainability: Its rigorous definition, major features and practical realisation

As mentioned above, the qualitatively big and nontrivial transition to the truly sustainable development way for the entire civilisation constitutes an integral part of the revolution of complexity in science and technology, implying also the social and intellectual transition to the superior level of consciousness. This result is very different from all usual, semi-empirical and basically wrong ideas about sustainability, as we show [55-57] that this unitary sustainability assumed without essential, qualitatively big transition to the complexity-growth regime in all human activities is strictly impossible and may actually play the negative role of a vain, but strongly dominating illusion, preventing the transition to genuine sustainability.

We show first [56] that the universal curve of complexity development from dynamic information to dynamic entropy (see Section 2.4) goes right now through a *major bifurcation point*, after which it can continue, within the current tendency and complexity level, only in the fatal complexity-destruction regime ending in the essential degradation down to qualitatively lower levels of complexity and consciousness, or else it can pass to the intrinsically creative and sustainable way after the transition to the complexity-creation mode at the superior level of consciousness and complexity development. It is important that this rigorously derived conclusion does not depend on the details of unitary regime realisation in the now dominating, default tendency, including the intensely imposed pseudo-"green" technologies and various "resource-saving" practices (they can only slightly slow down the occurring degradation, but also provide a dangerous illusion of problem solution until it will be too late for the genuine, complexity-driven sustainability transition providing the true solution). In other words, it is rigorously shown that we are in a special development point now, facing the "complexity barrier" after the global "complexity threshold", starting from which the traditional, "eternal" drive of purely empirical "invisible hand", in economy and elsewhere, becomes fundamentally inefficient for any further progress, forever and can now produce only increasing systemic degradation over all scales and dimensions (as opposed to the dominating idea of only a temporal crisis, however deep it may be, followed by the "natural", empirically driven new rise).

As to the concrete ways of realisation of the real sustainability transition and stable progress after that, they will always involve the key change from the traditional unitary complexity-destruction practices, approaches and thinking to the qualitatively different complexity-creation activities, in science, technology and engineering, social dynamics, settlement structure and related, unifying intellectual development [15,16,55-57]. This crucial and direct involvement of the unreduced dynamic complexity, rigorously and universally based on the causally complete, dynamically multivalued interaction problem solution (Section 2), constitutes the *well-defined, essential difference* from any dominating unitary ideas of sustainability and further civilisation progress.

We have demonstrated above the crucial advantages of the unreduced concept of dynamic complexity in understanding of intelligence and consciousness dynamics, in direct relation to urgent development problems (Sections 3 and 4). It is important that equally great, problem-solving advantages are provided by the same unreduced complexity analysis in *all other fields of fundamental science* (otherwise stagnating, or "ending"), from particle physics to efficient genetics, causal biology, integral medicine and creative ecology [15-24,55-57]. It is not difficult to see that the same is true for *applied science, technology and engineering development* and its accumulating unsolved, pressing problems related e.g. to the new, pure and practically unlimited energy sources, efficient, sustainable exploitation-development of all natural resources combined with growing infrastructure quality, or emerging medical and psychological difficulties [15,55-57]. The observed dangerous stagnation in these key directions, despite quantitatively huge and technically powerful efforts within the unitary approach, confirms the necessity of a qualitatively new vision, which we specify in the form of our unreduced dynamic complexity concept.

Similar general transformation from complexity-destruction to complexity-growth mode will inevitably take place in *social structure and science organisation and practice* itself, from the centralised and self-destructive Unitary system (the only type of organisation known until now, in various forms) to the distributed and intrinsically creative Harmonical system [15,55-57]. The new society of the harmonical level will be qualitatively different from any, even most "developed" unitary organisation, including the appearance of omnipresent *social consciousness* at the harmonical level located, of course, within each individual consciousness and definitely absent in only empirically, mechanistically driven unitary development mode. Such self-aware society is therefore able, in the normal way of its existence, to causally understand and guide its own development as that of a real complex system (with the above general purpose of further consciousness progress, Section 5.2), contrary to any unitary, even scientifically rich social organisation.

Needless to say, the organisation and role of the new, practically unlimited *knowledge creation system*, now inseparable from the entire social system and replacing the unitary science content and organisation, will also change and grow qualitatively after the sustainability transition, where "science" will not be isolated any more into an esoteric activity of self-estimated "sages", practically inaccessible for real "public understanding" (and thus any efficient control), but will instead be organically, inseparably interwoven with the fabric of entire social life and development as its major, driving element (which is evidently the only possibility for a society to be estimated as really "developed", but it can actually be realised only at the Harmonical level, after the complexity transition). That is why the sus-tainability and complexity transition can also be called the last scientific revolution [57] (with the reference to the famous analysis of Thomas Kuhn [68]), after which the development of science (and actually anything else) occurs in a stable and permanent way, without accumulating antagonistic contradictions followed by a characteristic disruptive "revolution" of unitary science (remaining thus a past phenomenon inherent only to that, very special, artificially limited kind of knowledge). It is evident that such new kind of progress of ever more *complex system* of self-aware knowledge-based society can be efficiently guided *only* by the causally complete content and dynamics of the *universal science of complexity*, such as the one tentatively outlined in previous sections, where the organisation and social involvement of that truly new kind of science forms itself a major integral part of the unified complex system of Harmonical, intrinsically sustainable society. The urgently needed *sustainability* can thus be *rigorously specified* itself as that superior level of consciousness and knowledge based on the unreduced and now properly *growing dynamic complexity* of the planetary environment consistently described in the above dynamic multivaluedness concept.